\documentclass[a4paper,fleqn,usenatbib,useAMS]{mnras}

\usepackage{graphicx}	
\usepackage{amsmath}	
\usepackage{amssymb}	
\usepackage{multicol}   
\usepackage{bm}		
\usepackage{pdflscape}	
\usepackage[export]{adjustbox}
\newcommand{\ct}{\citealt}
\renewcommand{\thesubsubsection}{\thesubsection.\alph{subsubsection}}

\usepackage[T1]{fontenc}
\usepackage{ae,aecompl}

\usepackage{newtxtext,newtxmath}

\title[Field Decoupling \& Disc Formation]{Decoupling of Magnetic Fields in Collapsing Protostellar Envelopes and Disc Formation and Fragmentation}

\author[B. Zhao et al.]{Bo Zhao$^{1}$\thanks{Contact e-mail: \href{mailto:bo.zhao@mpe.mpg.de}{bo.zhao@mpe.mpg.de}}\thanks{Present address: Giessenbachstr. 1, D-85748, Garching, Germany},
Paola Caselli$^{1}$,
Zhi-Yun Li$^{2}$,
Ruben Krasnopolsky$^{3}$\\
\\
$^{1}$Max-Planck-Institut f\"ur extraterrestrische Physik (MPE), Garching, Germany, 85748\\
$^{2}$University of Virginia, Astronomy Department, Charlottesville, USA, 22904\\
$^{3}$Academia Sinica Institute of Astronomy and Astrophysics, 10167, Taipei, Taiwan\\}

\pubyear{2017}

\begin{document}
\label{firstpage}
\pagerange{\pageref{firstpage}--\pageref{lastpage}}
\maketitle

\begin{abstract}
{Efficient magnetic braking is a formidable obstacle to the formation 
of rotationally supported discs (RSDs) around protostars in magnetized 
dense cores. We have previously shown, through 2D (axisymmetric) 
non-ideal MHD simulations, that removing very small grains (VSGs: 
$\sim$10~$\AA$ to few 100~$\AA$) can greatly enhance ambipolar diffusion 
and enable the formation of RSDs. Here we extend the simulations of 
disc formation enabled by VSG removal to 3D. We find that the key to 
this scenario of disc formation is that the drift velocity 
of the magnetic field almost cancels out the infall velocity of the neutrals 
in the $10^2$-$10^3$AU-scale ``pseudo-disc'' where the field lines are 
most severely pinched and most of protostellar envelope mass infall 
occurs. As a result, the bulk neutral envelope matter can collapse 
without dragging much magnetic flux into the disc-forming region, 
which lowers the magnetic braking efficiency. We find that the initial 
discs enabled by VSG removal tend to be Toomre-unstable, which leads 
to the formation of prominent spiral structures that function 
as centrifugal barriers. The piling-up of infall material near 
the centrifugal barrier often produces dense fragments of tens 
of Jupiter masses, especially in cores that are not too strongly 
magnetized. Some fragments accrete onto the central stellar 
object, producing bursts in mass accretion rate. Others are longer 
lived, although whether they can survive long-term to produce multiple 
systems remains to be ascertained. Our results highlight the importance 
of dust grain evolution in determining the formation and properties 
of protostellar discs and potentially multiple systems.}
\end{abstract}

\begin{keywords}
magnetic fields -MHD- circumstellar matter - stars: formation
\end{keywords}



\section{Introduction}
\label{Chap.Intro}

Angular momentum lies in the heart of the formation of rotationally 
supported discs (RSDs), which is strongly affected by the magnetic 
flux harboured in the central disc-forming region. 
The winding of pinched magnetic field lines by gas 
rotation can cause large magnetic torques that transport angular momentum 
away from the equatorial region \citep{MestelSpitzer1956,Mouschovias1977,MouschoviasPaleologou1980,BasuMouschovias1994}, 
preventing the formation of RSDs. This is the so-called magnetic braking 
problem \citep{Allen+2003,MellonLi2008,Li+2011}, which is 
in tension with recent high-resolution 
observations of $\sim$100~AU discs and spirals around young protostellar 
objects \citep{Tobin+2012,Tobin+2013,Tokuda+2014,Perez+2016,Tobin+2016}.

Recent theoretical studies on disc formation often resort to 
non-ideal MHD effects to avert the magnetic braking ``catastrophe'' 
\citep{Konigl1987,MouschoviasPaleologou1986,MellonLi2009,DappBasu2010,Li+2011,Krasnopolsky+2011,Dapp+2012,Tomida+2013,Tomida+2015,Tsukamoto+2015a,Tsukamoto+2015b,Masson+2015,Wurster+2016}, 
based on the principle that magnetic fields are partially 
decoupled from neutral matter in weakly ionized dense cores 
\citep{BerginTafalla2007}. However, most studies focus on the decoupling 
of the magnetic field at very high densities within a few to tens of AU scale 
around the central star. The RSDs formed in this fashion, if any, are 
of limited size (at most $\sim$30~AU in radius) and have no clear sign 
of growth to the $\sim$50-100~AU Keplerian discs observed around 
Class 0 protostars \citep{Murillo+2013,Codella+2014,Tobin+2016,Lee+2017}.

In our previous two-dimensional (2D) study, \citet{Zhao+2016} 
(Zhao+16 hereafter), we have shown that large (axisymmetric) 
RSDs and self-gravitating rings are able to form in the absence of 
very small grains (VSGs) of $\sim$10~$\AA$ to few 100~$\AA$. 
In fact, it is the large population of VSGs in the MRN size 
distribution that dominates the coupling of the bulk neutral matter to 
the magnetic field at densities below 10$^{10}$~cm$^{-3}$. The removal 
of VSGs can enhance the ambipolar diffusion of magnetic field in the 
envelope by $\sim$1--2 orders of magnitude, so that the amount of 
magnetic flux dragged by the collapse into the central disc-forming region 
is reduced. Because the decoupling of magnetic fields occurs much earlier, 
magnetic braking operating in the central equatorial region becomes 
inefficient in transporting away angular momentum. Therefore, the 
high specific angular momentum in the infalling gas eventually leads 
to the formation and growth of early RSDs.

According to Zhao+16, the key requirement for magnetized disc formation 
is the lack of VSGs in dense cores, which has been confirmed by 
recent CARMA centimetre survey searching for emission of spinning 
dust grains \citep{Tibbs+2016}. They show a depletion of nanometre grains 
($\lesssim 100~\AA$) in all dense molecular cores in their survey. 
The depletion of VSGs can occur either in the prestellar or protostellar 
phase, through both accretion of VSGs onto dust grain mantles 
\citep[process analogous to molecular freeze-out, e.g.,][]{TielensHagen1982,Hasegawa+1992} 
and grain coagulation \citep[e.g.,][]{Chokshi+1993,DominikTielens1997}. 
Grain coagulation has been shown to be rather efficient in removing 
small grains (<0.1~$\mu$m) within a few 10$^6$ years 
\citep{Ossenkopf1993,Hirashita2012}. 
Furthermore, the timescale for coagulation of VSGs onto big grains has 
shown to be the shortest \citep[$\sim$1.6$\times$10$^3$~years;][]{Kohler+2012}.
Hence, prior to the birth of the protostellar disc, the collapsing envelope 
is likely to have a greatly reduced abundance of VSGs, 
which enhances ambipolar diffusion.

In this study, we extend our previous study of AD-induced disc 
formation to three-dimension (3D) and provide conditions for the 
formation of RSDs and multiple systems.
We confirm the crucial role of removing VSGs in the formation of RSDs, as 
found in Zhao+16. Particularly, we find that the infall velocity of 
the magnetic field decreases to nearly zero over a wide equatorial 
region in the envelope, 
where magnetic field lines are strongly pinched by the collapsing flow. 
Unlike \citet{Hennebelle+2016}, the self-regulation of magnetic fields 
by AD actually occurs on much larger scales, and is due to the 
vertical gradient of the radial magnetic field 
(field pinching and the associated magnetic tension force) 
instead of the radial gradient of poloidal fields (and the associated 
magnetic pressure gradient). Efficient decoupling of magnetic fields 
in the envelope leads to sufficient angular momentum 
influx into the disc forming region, triggering the formation of 
an early RSD or ring ($\sim$10-40~AU) which evolves into a small 
circumstellar disc (~$\sim$20~AU) surrounded by large spiral 
structures (a few 100~AU). We also confirm that both rotation speed 
and magnetic field strength of the initial core can affect 
the morphology and size of the disc. For cores with lower magnetization, 
multiple companion objects can form from spiral or ring structures 
via material piling up near the centrifugal barrier, which belongs to 
the type of fragmentation driven by rapid accretion 
\citep{Kratter+2010,KratterLodato2016}. 
The models that produce large spirals and multiple systems in 3D 
correspond to the models with ring structures found in the 2D 
axisymmetric calculations of Zhao+16. 

The rest of the paper is organized as follows. 
Section \ref{Chap.IC} describes the 
initial conditions of the simulation set, together with an overview of 
the results. In Section \ref{Chap.SimulResult}, we present and analyse 
the simulation results. The comparison with existing theories of disc 
formation and a case study of B335 are given in Section 
\ref{Chap.Discuss}. Finally, we summarize the results in Section 
\ref{Chap.Summary}.

\section{Initial Condition}
\label{Chap.IC}

We carry out three-dimensional (3D) numerical simulations\footnote{
Merely for convenience, we follow the axisymmetric prestellar collapse 
in 2D until well before the formation of a first hydrostatic core 
\citep{Larson1969}, and restart the simulation in full 3D thereafter.}
using ZeusTW code \citep{Krasnopolsky+2010}, focusing on the 
ambipolar diffusion for the diffusion of magnetic field in collapsing cloud
cores. We adopt the same chemical network as in \citet{Zhao+2016}, which 
computes the magnetic diffusivities\footnote{The momentum transfer 
rate coefficients are parametrized as a function of temperature according 
to Table 1 of \citet{PintoGalli2008}.} at every hydrodynamic timestep 
and at each spatial point. 

The initial conditions are similar to \citet{Zhao+2016}, except for 
a more accurate equation of state that is described in Appendix~\ref{App.A}. 
We initialize a uniform, isolated spherical core with total mass 
$M_{\rm c}=1.0~M_{\sun}$, and radius $R_{\rm c}=10^{17}$~cm~$\approx 6684$~AU. 
This corresponds to an initial mass density 
$\rho_0=4.77 \times 10^{-19}$~g~cm$^{-3}$ and a number density for 
molecular hydrogen $n({\rm H}_2)=1.2 \times 10^5$~cm$^{-3}$ 
(assuming mean molecular weight $\mu=2.36$). The free-fall time of the core 
is thus $t_{\rm ff} =3 \times 10^{12}$~s~$\approx 9.6 \times 10^4$~yr. 
The initial core is rotating as a solid-body with angular speed 
$\omega_0=1 \times 10^{-13}$~s$^{-1}$ for slow rotating case, and 
$2 \times 10^{-13}$~s$^{-1}$ for fast rotating case, corresponding to 
a ratio of rotational to gravitational energy $\beta_{\rm rot}\approx$ 
0.025 and 0.1 \citep{Goodman+1993}, respectively. 
The initial core is threaded by a uniform magnetic field along the 
rotation axis with a constant strength $B_0$ of 
$4.25 \times 10^{-5}$~G for strong field case and 
$2.13 \times 10^{-5}$~G for weak field case, which gives a 
dimensionless mass-to-flux ratio $\lambda$ 
($\equiv {M_{\rm c} \over \pi R_{\rm c}^2 B_0}2\pi\sqrt{G}$) 
of 2.4 and 4.8, respectively. 
It is consistent with the mean value of $\lambda$ inferred from the 
OH Zeeman observations by \citet{TrolandCrutcher2008}.

We adopt the spherical coordinate system ($r$, $\theta$, $\phi$) and 
non-uniform grid to provide high resolution towards the innermost 
region of simulation domain. 
The inner boundary has a radius $r_{\rm in}=3 \times 10^{13}$~cm~$=2$~AU 
and the outer has $r_{\rm out}=10^{\rm 17}$~cm. At both boundaries, we 
impose a standard outflow boundary conditions to allow matter to leave 
the computational domain. The mass accreted across the inner boundary 
is collected at the centre as the stellar object. We use a total of 
$120 \times 96 \times 96$ grid points. The grid is uniform in the 
$\phi$-direction, non-uniform in the $\theta$-direction with 
$\delta \theta=0.6713^{\circ}$ near the equator, and non-uniform in the 
$r$-direction with a spacing $\delta r=0.1$~AU next to the inner boundary. 
The $r$-direction spacing increases geometrically outward 
by a constant factor of $\sim$1.0733, and the $\theta$-direction 
spacing increases geometrically from the equator to either pole by a 
constant factor of $\sim$1.0387.

In most models of this study, we fix the cosmic-ray ionization rate 
at the cloud edge to be $\zeta_0^{\rm H_2}=1.0\times 10^{-17}$~s$^{-1}$, 
with a characteristic attenuation length of 96~g~cm$^{-2}$. 
We do not consider the high cosmic-ray ionization case of 
$\zeta_0^{\rm H_2}=5.0\times 10^{-17}$~s$^{-1}$ of Zhao+16, where 
disc formation is strongly suppressed unless both $\beta_{\rm rot}$ 
and $\lambda$ are high. We choose two grain size distributions, MRN and 
tr-MRN for the computation of non-ideal MHD diffusivities. Both size 
distributions have the same power law index $-3.5$ and maximum grain size 
$a_{\rm max}$=0.25~$\mu$m, but different minimum grain sizes 
$a_{\rm min}$=0.005~$\mu$m (MRN) and 0.1~$\mu$m (tr-MRN). As compared 
to Zhao+16, we dropped the large grain (LG) case, as we find its 
effect on disc formation is in between the MRN and tr-MRN cases. 
The simulation models are summarized in Table~\ref{Tab:models}. 
\begin{table*}
\begin{minipage}{150mm}
\caption{Model Parameters \& Result Summary}
\label{Tab:models}
\begin{tabular}{lccllccc}
\hline\hline
Model & Grain Size & $\lambda$ & $\beta_{\rm rot}$ & Morphology & 
Initial Disc/Ring & Spiral Structure & Circumstellar Disc \\
 & & & & & Radius (AU) & Radius (AU) & Average Radius (AU) \\
\hline
2.4Slw-MRN & MRN & 2.4 & 0.025 & DEMS & -- & -- & -- \\
2.4Slw-trMRN & tr-MRN & 2.4 & 0.025 & Disc+Spiral & $\sim$15 & $\sim$40 & $\sim$20 \\
2.4Fst-MRN & MRN & 2.4 & 0.1 & DEMS & -- & -- & -- \\
2.4Fst-trMRN & tr-MRN & 2.4 & 0.1 & Ring$\rightarrow$Disc+Spiral & $\sim$30 & $\sim$100$\uparrow$ & $\sim$15 \\
\hline
4.8Slw-MRN & MRN & 4.8 & 0.025 & DEMS+Disc$^{\rm Trans}$ & $\sim$15 & -- & -- \\
4.8Slw-trMRN & tr-MRN & 4.8 & 0.025 & Ring$\rightarrow$Disc+Spiral$^{\rm Frag}$ & $\sim$25 & $\sim$150$\uparrow$ & $\sim$20 \\
4.8Fst-MRN & MRN & 4.8 & 0.1 & Disc$^{\rm Shrink}$$\rightarrow$DEMS & $\sim$20 & -- & $\sim$10$\rightarrow$0 \\
4.8Fst-trMRN & tr-MRN & 4.8 & 0.1 & Ring$^{\rm Frag}$ & $\sim$40 & $\sim$200$\uparrow$ & $\sim$10 \\
\hline
HydroSlw & -- & $\infty$ & 0.025 & Ring$^{\rm Frag}$$\rightarrow$Multiples & $\sim$50--100 & $\sim$500$\uparrow$ & $\sim$20 \\
HydroFst & -- & $\infty$ & 0.1 & Ring$^{\rm Frag}$$\rightarrow$Multiples & $\sim$70--150 & $\sim$1000$\uparrow$ & $\sim$20 \\
\hline
\end{tabular}

$\dagger$~MRN: full MRN distribution with $a_{\rm min}=0.005~\mu$m, 
$a_{\rm max}=0.25~\mu$m \\
$\dagger$~tr-MRN: truncated MRN with $a_{\rm min}=0.1~\mu$m, 
$a_{\rm max}=0.25~\mu$m \\
$\dagger$~DEMS: Decoupling-Enabled Magnetic Structures \citep{Zhao+2011}. \\
$\dagger$~RSD$^{\rm Trans}$: a transient RSD appears early on 
due to the high angular momentum initially, but disappears within 10$^3$~yr. \\
$\dagger$~RSD$^{\rm Shrink}$: initially forms a RSD but shrinks in size 
over time ($\sim$10$^3$~yr). \\
$\dagger$~Spiral$^{\rm Frag}$ or Ring$^{\rm Frag}$: fragmentation occurs 
in spirals or rings, leading to the formation of binary and multiple systems. \\
$\dagger$~$\uparrow$: the radius of the spiral structure (or centrifugal radius) continues to increase over time.
\\
\end{minipage}
\end{table*}

\section{Simulation Results}
\label{Chap.SimulResult}

As shown in Table~\ref{Tab:models}, the conditions for disc formation are 
largely similar to Zhao+2016, with formation of rotational supported 
structures occurring in all tr-MRN cases, 
compared to no disc, transient disc, or shrinking disc in the MRN cases. 
The main difference from Zhao+16 is the variety of disc morphologies 
shown in 3D, ranging from DEMS 
\citep[Decoupling-Enabled Magnetic Structures;][]{Zhao+2011} to RSDs, 
and from spiral to ring structures. 
In cases with lower magnetization ($\lambda=4.8$), fragmentation often 
occurs within spirals and rings that leads to the initial formation of 
binary and multiple systems.

\subsection{AD-Enabled Disc Formation: Field Decoupling in the Envelope}
\label{S.AD-Decouple}

Despite existing literature on field decoupling in the 
high density disc itself, we find the key for disc 
formation lies in the field decoupling in the protostellar envelope. 
Particularly, an enhanced AD in the absence of VSGs (Zhao+16) 
ensures sufficient magnetic flux to be decoupled in the low-density 
envelope before reaching the inner tens-of-AU stellar vicinity, thus 
preventing the ``catastrophic'' magnetic braking and preserving 
sufficient angular momentum for disc assembly.

\subsubsection{Truncated tr-MRN Model: 2.4Slw-trMRN}
\label{S.model-trMRN}

\begin{figure*}
\includegraphics[width=\textwidth]{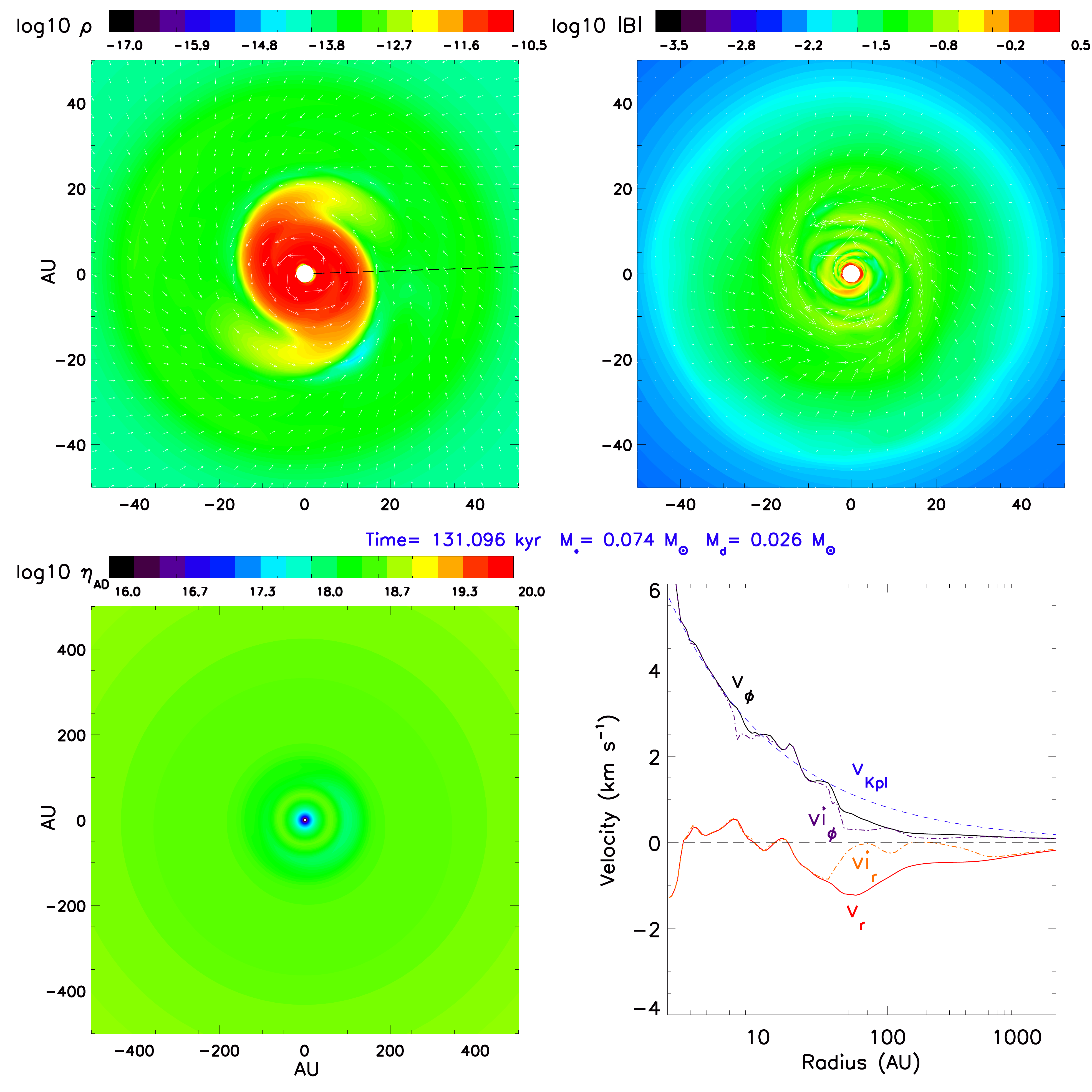}
\caption{Distributions of logarithmic mass density $\rho$ (g~cm$^{-3}$), 
magnetic field strength |$B$| (G), and ambipolar diffusivity 
$\eta_{\rm AD}$ (cm$^2$~s$^{-1}$) within 0.336$^{\circ}$ of the equator, 
and velocity profile along the $\phi=1.9^{\circ}$ line cut 
(black dashed line in the top-left panel) for the 2.4Slw-trMRN model 
at time $t \approx 131.1$~kyr$=1.36t_{\rm ff}$. 
The projected velocity field (top-left) and magnetic field (top-right) 
are shown as white arrows. Length unit of the axes is in AU. 
The typical temperature in the disc is around 100~K.}
\label{Fig:2.4Slw-trMRN_D}
\end{figure*}
\begin{figure}
\includegraphics[width=\columnwidth]{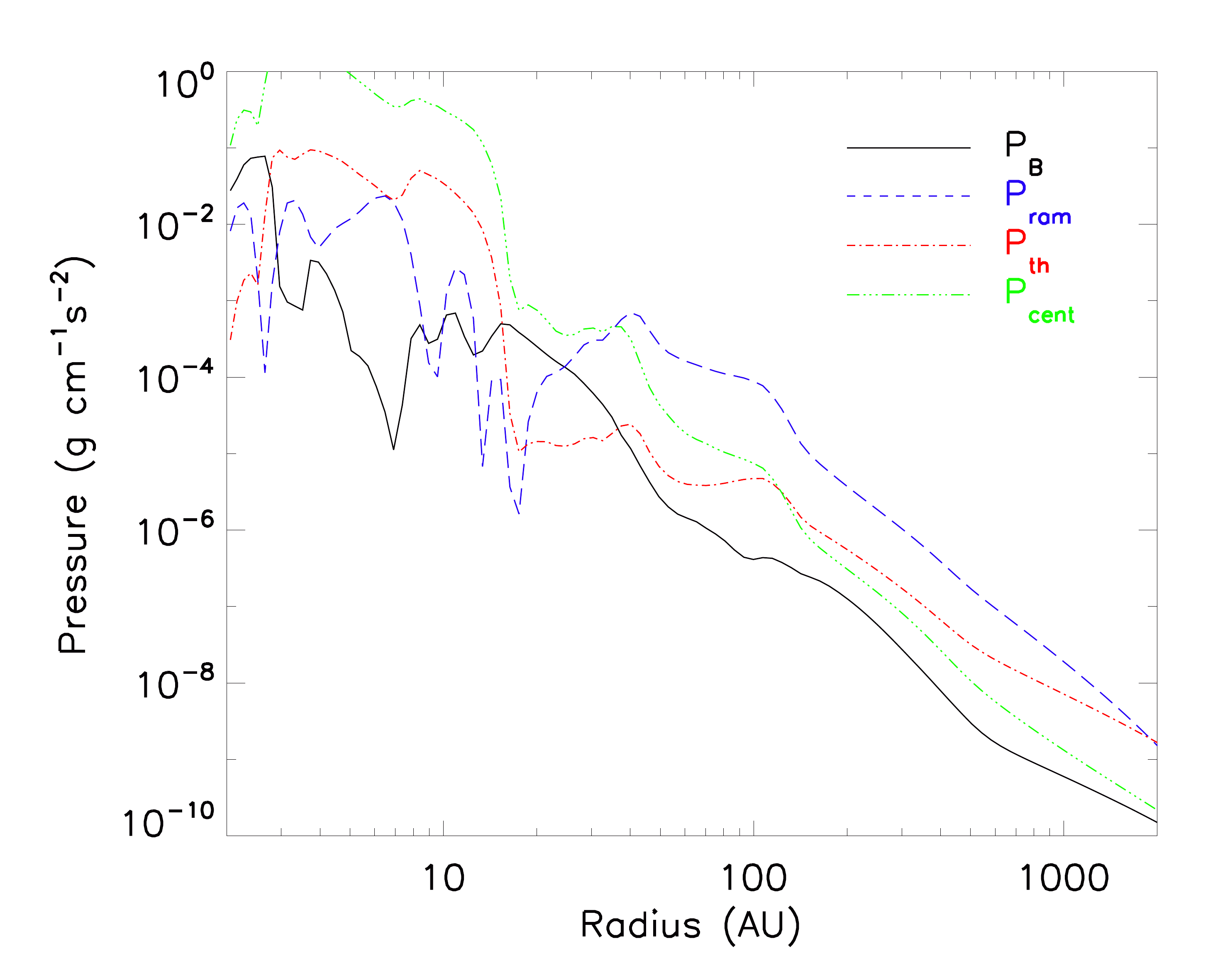}
\caption{Profile of thermal $P_{\rm th}$, magnetic $P_{\rm B}$, 
ram $P_{\rm ram}$, and effective centrifugal $P_{\rm cent}$ pressures 
within 0.336$^{\circ}$ of the equator, along the same line 
cut as the velocity profile in Fig.~\ref{Fig:2.4Slw-trMRN_D}.}
\label{Fig:Disc_Pres}
\end{figure}

Fig.~\ref{Fig:2.4Slw-trMRN_D} shows the face-on view of the equatorial 
disc-spiral structure in the 2.4Slw-trMRN model at 131.096~kyr, 
about 4.56~kyr after the formation of first core. The velocity profile 
clearly shows that the inner $\sim$20~AU disc is rotating with 
Keplerian speed. The spiral structure is likely infall streams 
onto the Keplerian disc. The effective centrifugal pressure 
$P_{\rm cent}={1 \over 2} \rho {\rm v}_\phi^2$ in the disc 
is $\sim$10 times higher than the thermal pressure, indicating that 
the disc is primarily rotationally supported (see Fig.~\ref{Fig:Disc_Pres}). 
The plasma-$\beta$ ($\equiv {P_{\rm th} \over P_{\rm B}}$, where 
$P_{\rm th}$ is the thermal pressure and $P_{\rm B}$ the magnetic 
pressure) in the disc is on the order of $\sim$10$^2$. 
The disc rotation also wraps up magnetic field lines over time, 
producing the well-developed toroidal magnetic field components shown 
in the top-right panel of Fig.~\ref{Fig:2.4Slw-trMRN_D}.

The direct evidence of magnetic decoupling in the envelope is 
the significant separation between the infall velocity of neutral gas 
v$_{\rm r}$ and that of the ions vi$_{\rm r}$ (bottom-right panel of 
Fig.~\ref{Fig:2.4Slw-trMRN_D}). 
The ion velocity is in fact an ``effective ion velocity'' defined as 
$\bmath{\rm vi} \equiv \bmath{\rm v}+\bmath{\rm v}_{\rm d}$, 
where $\bmath{\rm v}_{\rm d}$ is the drift velocity of the magnetic field 
given in Eq.~\ref{Eq:vdr}. Because AD in zeusTW is treated using 
an explicit method \citep{MacLow+1995,Li+2011} and $\bmath{\rm vi}$ 
is directly used to evolve the magnetic field, hence $\bmath{\rm vi}$ 
actually denotes the velocity of the magnetic field.\footnote{In reality, 
different charged species are attached to the magnetic field to different 
degrees, and the velocity of individual charged species 
can differ from the magnetic field velocity. 
We refer the readers to \citet{KunzMouschovias2009,KunzMouschovias2010} 
for detailed discussions.}
The nearly zero effective ion velocity\footnote{The velocity profile is plotted 
along a line cut within 0.336$^{\circ}$ of the equator; 
however, the infall stream in this model does not lie on 
the equator and lands upon the disc from the upper hemisphere 
(see \S~\ref{S.fieldPinch} and Fig.~\ref{Fig:ad_Vel}) Therefore, 
the actual decoupling region along the infall stream should be slightly 
bigger and a moderate decoupling indeed occurs near 20~AU as well.}
vi$_{\rm r}$ at $\sim$50~AU--300~AU implies that magnetic fields 
have almost decoupled from the bulk infall 
motion of neutral gas and are left behind in the envelope. Such an 
efficient decoupling is a result of the enhanced AD at low density 
regimes ($\lesssim10^{10}$~cm$^{-2}$) in the absence of VSGs (Zhao+16; 
see their Section 4 and Fig.~2). It is clearly manifested in 
the bottom-left panel of Fig.~\ref{Fig:2.4Slw-trMRN_D} that 
the AD diffusivity on the $\sim$500~AU scale is already well-above 
10$^{18}$~cm$^2$~s$^{-1}$, which is at least one order of magnitude 
higher than the infalling region in the 2.4Slw-MRN model (shown next).

\subsubsection{MRN Grain Model: 2.4Slw-MRN}
\label{S.model-MRN}

\begin{figure*}
\includegraphics[width=\textwidth]{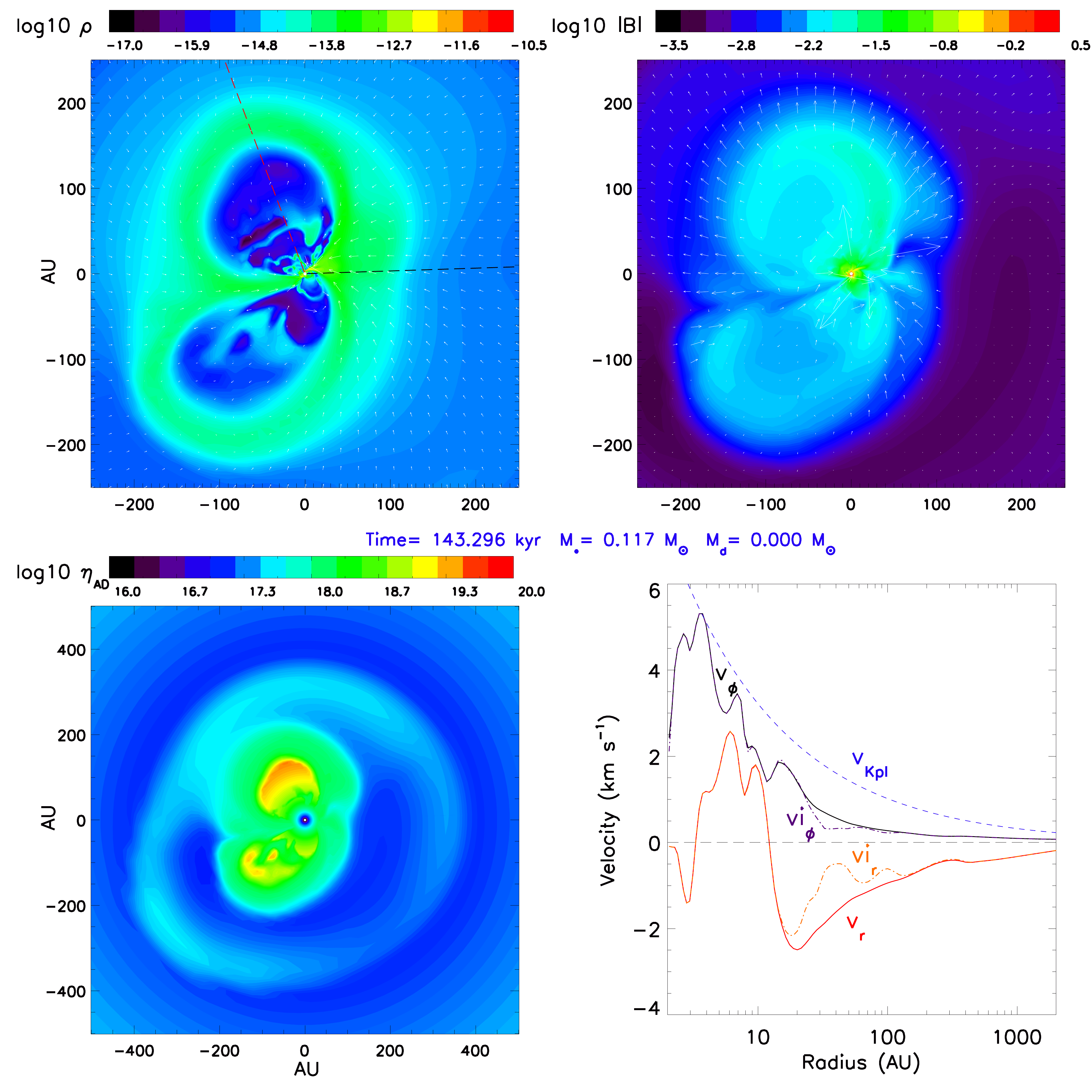}
\caption{Distributions of logarithmic mass density $\rho$ (g~cm$^{-3}$), 
magnetic field strength |$B$| (G), and ambipolar diffusivity 
$\eta_{\rm AD}$ (cm$^2$~s$^{-1}$) within 0.336$^{\circ}$ of the equator, 
and velocity profile along the $\phi=1.9^{\circ}$ line 
cut (black dashed line in the top-left panel) for the 2.4Slw-MRN model 
at time $t \approx 143.3$~kyr~$=1.49t_{\rm ff}$. 
The projected velocity field (top-left) and magnetic field (top-right) 
are shown as white arrows. Length unit of the axes is in AU. 
The red dashed line in the top-left panel is the line cut used 
in Fig.~\ref{Fig:DEMS_Anly}.}
\label{Fig:2.4Slw-MRN_D}
\end{figure*}

In the 2.4Slw-MRN model, the physical structure of the central region is 
drastically different. No obvious RSD forms at all 
(Fig.~\ref{Fig:2.4Slw-MRN_D}); instead, large torus-shaped DEMS occupy 
the inner $\sim$150~AU, which consist essentially of the decoupled flux 
from the accreted matter \citep{Zhao+2011,Krasnopolsky+2012}. 
Accordingly, the magnetic field strength is also higher in the 
low density DEMS. Such a structure expands over time from 
a few AU to a few 100~AU, which blocks infall and obstructs rotation 
over a large region on the equator. Gas finds other paths to reach the 
centre, however, dragging in most magnetic field lines as well 
(bottom right panel of Fig.~\ref{Fig:2.4Slw-MRN_D}, 
plotted along a line cut in the positive x-direction). 
Unlike the 2.4Slw-trMRN model, the effective ion velocity is almost 
indistinguishable from the bulk neutral velocity beyond 70~AU; 
There is only a partial separation between 
vi$_{\rm r}$ and v$_{\rm r}$ in the inner $\sim$20-70~AU; yet 
the separation is not as large as in the 2.4Slw-trMRN model. 
As most of the magnetic field is dragged all the way to the centre 
and eventually decouples at the inner boundary ($\sim$2~AU) 
from the accreted matter that enters the sink hole\footnote{When 
magnetic field lines are dragged to the inner boundary, they are 
free to move around after detaching from the accreted matter. 
Therefore, the total magnetic flux is conserved 
\citep[unlike][]{Wurster+2017}.}
it joins the existing DEMS or creates new DEMS in directions of 
least resistance. 

\begin{figure*}
\begin{tabular}{ll}
\includegraphics[width=\columnwidth]{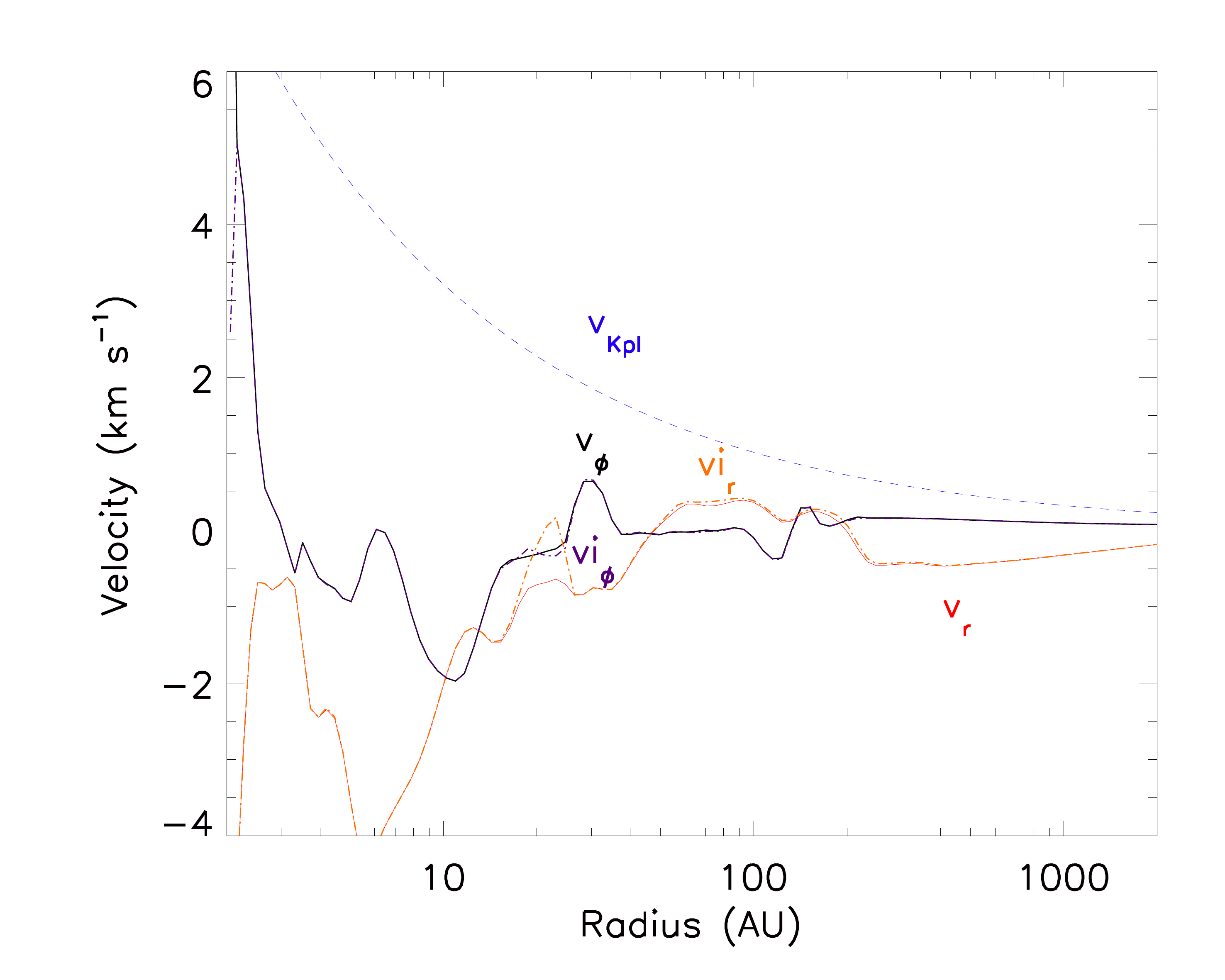}
\includegraphics[width=\columnwidth]{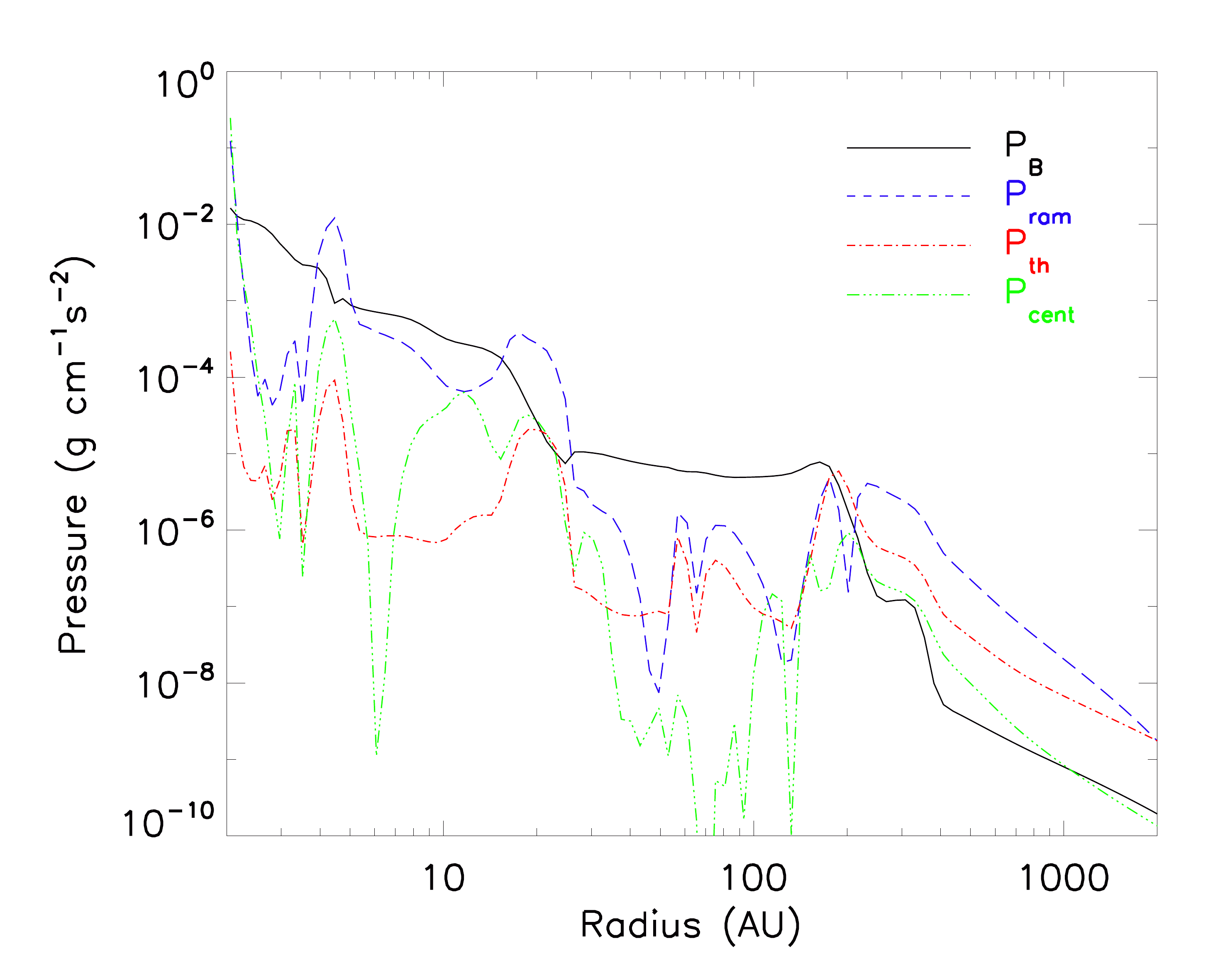}
\end{tabular}
\caption{Left panel: profile of equatorial infall and rotation speed in 
the 2.4Slw-MRN model at $t \approx 143.3$~kyr 
(same time as in Fig.~\ref{Fig:2.4Slw-MRN_D}). 
The Keplerian speed is plotted based on the central mass. 
Right panel: profile of equatorial thermal $P_{\rm th}$, magnetic $P_{\rm B}$, 
ram $P_{\rm ram}$, and effective centrifugal $P_{\rm cent}$ pressures. 
Both panels are plotted along a line cut through the DEMS with 
$\phi=111^{\circ}$ (red dashed line in the top-left panel of 
Fig.~\ref{Fig:2.4Slw-MRN_D}).}
\label{Fig:DEMS_Anly}
\end{figure*}

As shown in Fig.~\ref{Fig:DEMS_Anly}, the infall velocity 
for the northern DEMS component shows clear expansion 
with ${\rm v_r}$ and ${\rm vi_r}$ being positive between $\sim$50--200~AU. 
Very little rotation is present across the DEMS. The equatorial region 
within $\sim$30--200~AU is dominated by magnetic pressure $P_{\rm B}$; 
and the innermost $\sim$30~AU is dominated by both magnetic and ram 
pressures, which implies magnetic instability of the interchange type 
\citep[rising of more strongly magnetized material against 
less strongly magnetized collapsing flow, e.g.,][]{Parker1979}

In comparison to the 2.4Slw-trMRN model above, the primary reason for 
the failure of disc formation and the presence of DEMS is that too much 
magnetic flux has been brought to the centre. It is a direct 
consequence of the low ambipolar diffusivity in the infalling envelope 
beyond $\sim$10$^2$~AU 
(number density $\sim$10$^7$~g~cm$^{-3}$--10$^{10}$~g~cm$^{-3}$).
Apart from the expanding DEMS, the $\eta_{\rm AD}$ is mostly around 
10$^{17}$~cm$^2$~s$^{-1}$ within the 500~AU region (bottom left 
panel of Fig.~\ref{Fig:2.4Slw-MRN_D}) --- at least one 
order of magnitude lower than the values in the 2.4Slw-trMRN model. 
This leads to very little decoupling of magnetic field from the 
infall flow, and hence an excessive amount of magnetic flux reaching 
the centre. Although a certain degree of field decoupling indeed occurs 
in the inner few tens of AU, it does not prevent enough magnetic flux 
from accumulating in the central region and hence is unable to save the disc.
The drastic difference between the two models, is simply caused 
by changing a single parameter -- the grain size distribution. 

\subsubsection{Collapse Induced Pinching of B-Fields \\
\& B-Field Decoupling in the Envelope}
\label{S.fieldPinch}

\begin{figure*}
\includegraphics[width=1.03\textwidth]{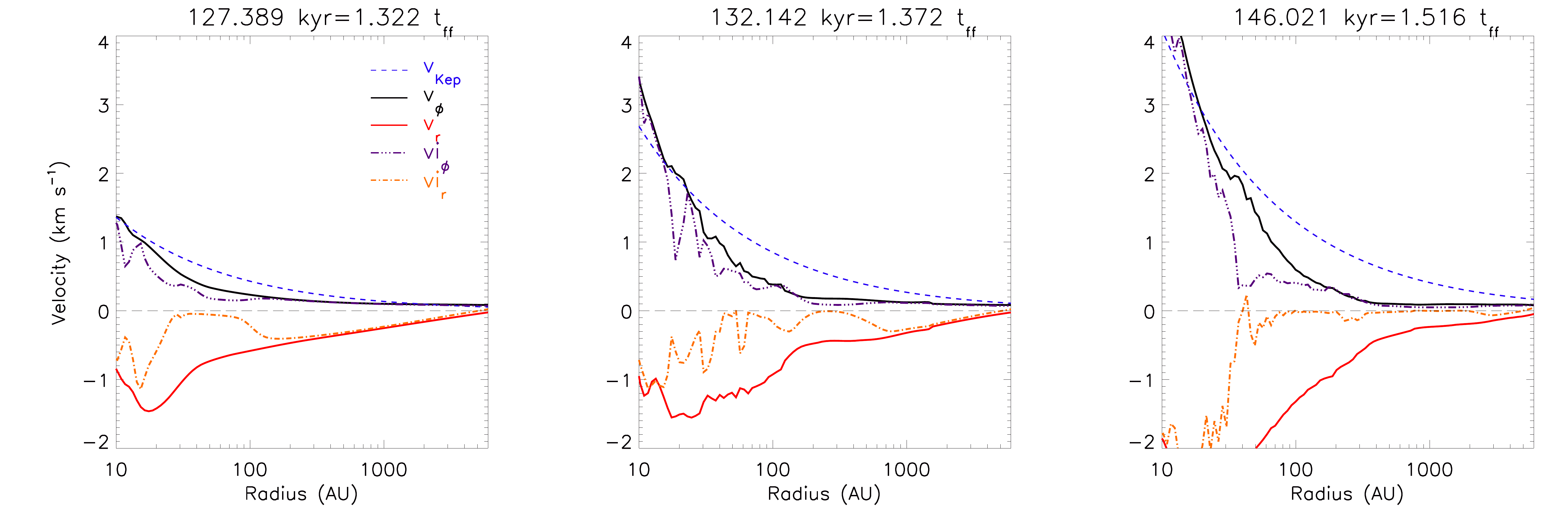}
\includegraphics[width=0.97\textwidth]{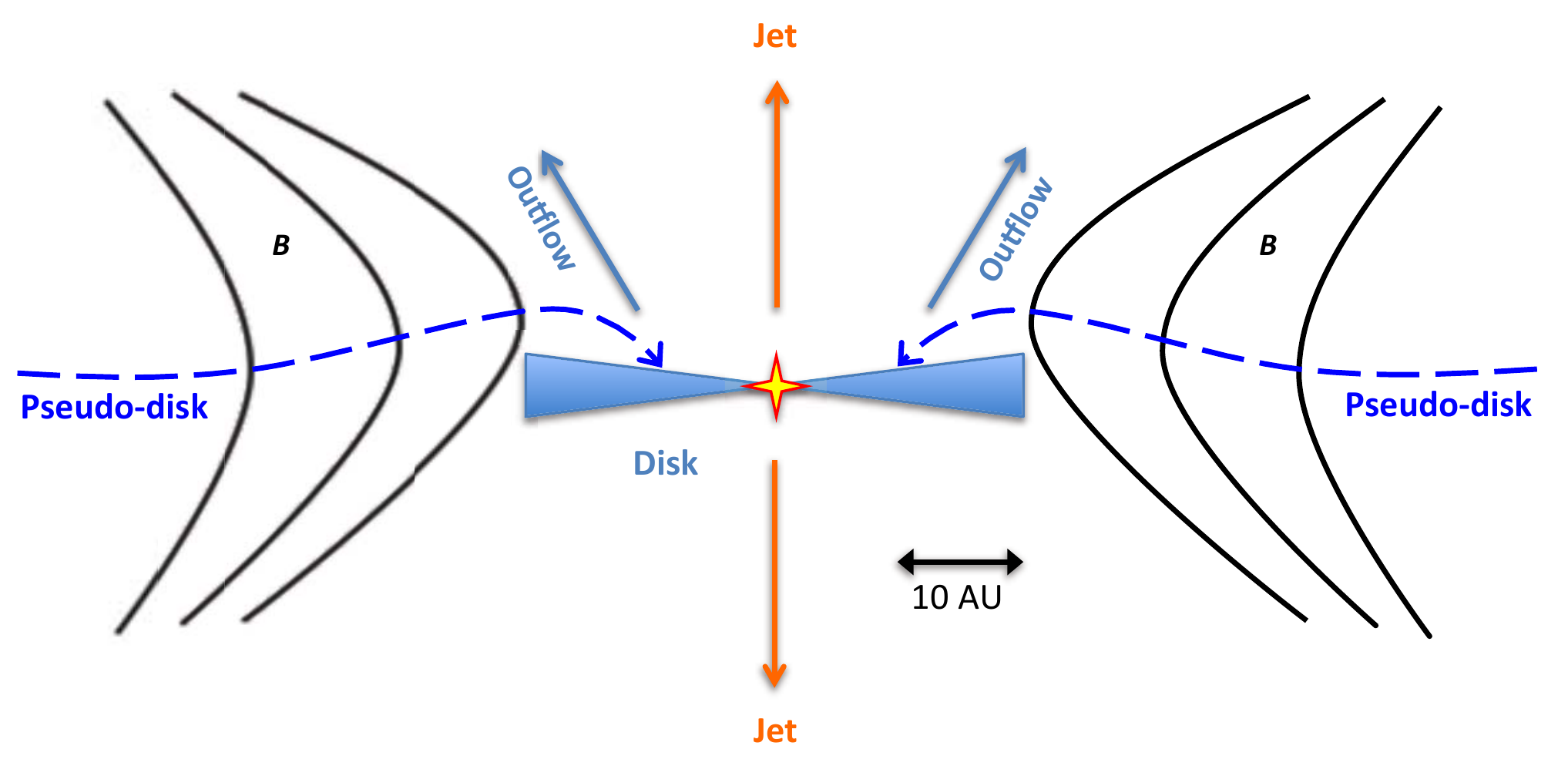}
\caption{Top row: velocity profiles for the 2.4Slw-trMRN model at time 
$t$=1.322~$t_{\rm ff}$, 1.372~$t_{\rm ff}$, and 1.516~$t_{\rm ff}$, 
along the curved infall stream or the pseudo-disc where field pinching 
are the strongest. 
Bottom row: illustration of the equatorial RSD and the curved pseudo-disc 
for the 2.4Slw-trMRN model (see also Fig.~\ref{Fig:Outflow_comp}). Such 
a configuration is persistent rather than transient.}
\label{Fig:ad_Vel}
\end{figure*}

In the tr-MRN cases where VSGs are absent, the regions of field 
decoupling expand over time from a few tens of AU to a few 10$^3$~AU 
into the envelope. As shown in Fig.~\ref{Fig:ad_Vel}, the effective infall 
velocity of ions ${\rm vi_r}$ is close to 0 between 30--80~AU at an early time 
$t \approx 127.4$~kyr ($\sim$850~yr after the formation of the first core), 
along with a well-decoupled zone between 20--100~AU. 
At $t \approx 132.1$~kyr, regions of zero ${\rm vi_r}$ are located 
between 40--400~AU, and the decoupling region extends further to 20--600~AU. 
At the later time $t \approx 146.0$~kyr, ${\rm vi_r}$ has reached 0 
in a wide equatorial region between 70--2000~AU, and is mostly 
separated from the neutral infall velocity ${\rm v_r}$ between 20--2000~AU. 
Note that the effective rotational velocity of ions ${\rm vi}_\phi$ 
is mostly identical to the neutral ${\rm v}_\phi$, with a moderate 
separation in the inner $\sim$100~AU, where poloidal magnetic fields 
start to wind up to generate the toroidal field components and a 
corresponding radial currents across the ``pseudo-disc'' 
\citep{GalliShu1993}.

The large separation between ${\rm vi_r}$ and ${\rm v_r}$ in the tr-MRN case 
indicates a notable radial drift of magnetic field relative to the neutrals, 
i.e., field decoupling from the collapsing flow. We show below that the 
location of the decoupling region is tied closely to the regions with strong 
pinching of magnetic field by the collapsing flow. 
Therefore, the expansion of the decoupling region is 
a natural consequence of the inside-out collapse that gradually induces 
the field pinching at larger and larger radii along the equator. 

The drift velocity $\bmath{\rm v}_{\rm d}$ of the magnetic field 
relative to the neutrals due to AD is defined as, 
\begin{equation}
\label{Eq:vdr}
\bmath{\rm v}_{\rm d}={\eta_{\rm AD} \over \bmath{B}^2} (\nabla \times \bmath{B}) \times \bmath{B}~.
\end{equation}
In a spherical coordinate system, the $r$-direction component of the drift 
velocity can be written as, 
\begin{equation}
\label{Eq:vdr_deriv}
\begin{split}
{\rm v}_{{\rm d},r} & = {\eta_{\rm AD} \over \bmath{B}^2} \left[ (\nabla \times \bmath{B})_\theta B_\phi - (\nabla \times \bmath{B})_\phi B_\theta \right] \\
& \approx -{\eta_{\rm AD} \over \bmath{B}^2}{B_\theta \over r} \left[ {\partial \over \partial r}(r B_\theta) - {\partial B_r \over \partial \theta} \right]~,
\end{split}
\end{equation}
where we ignore the contribution of $(\nabla \times \bmath{B}_\theta) B_\phi$ 
because in the collapsing envelope both the magnetic current in the 
$\theta$-direction and the field strength in the $\phi$-direction 
(toroidal field) are much smaller than the other corresponding components. 

\begin{figure}
\includegraphics[width=\columnwidth]{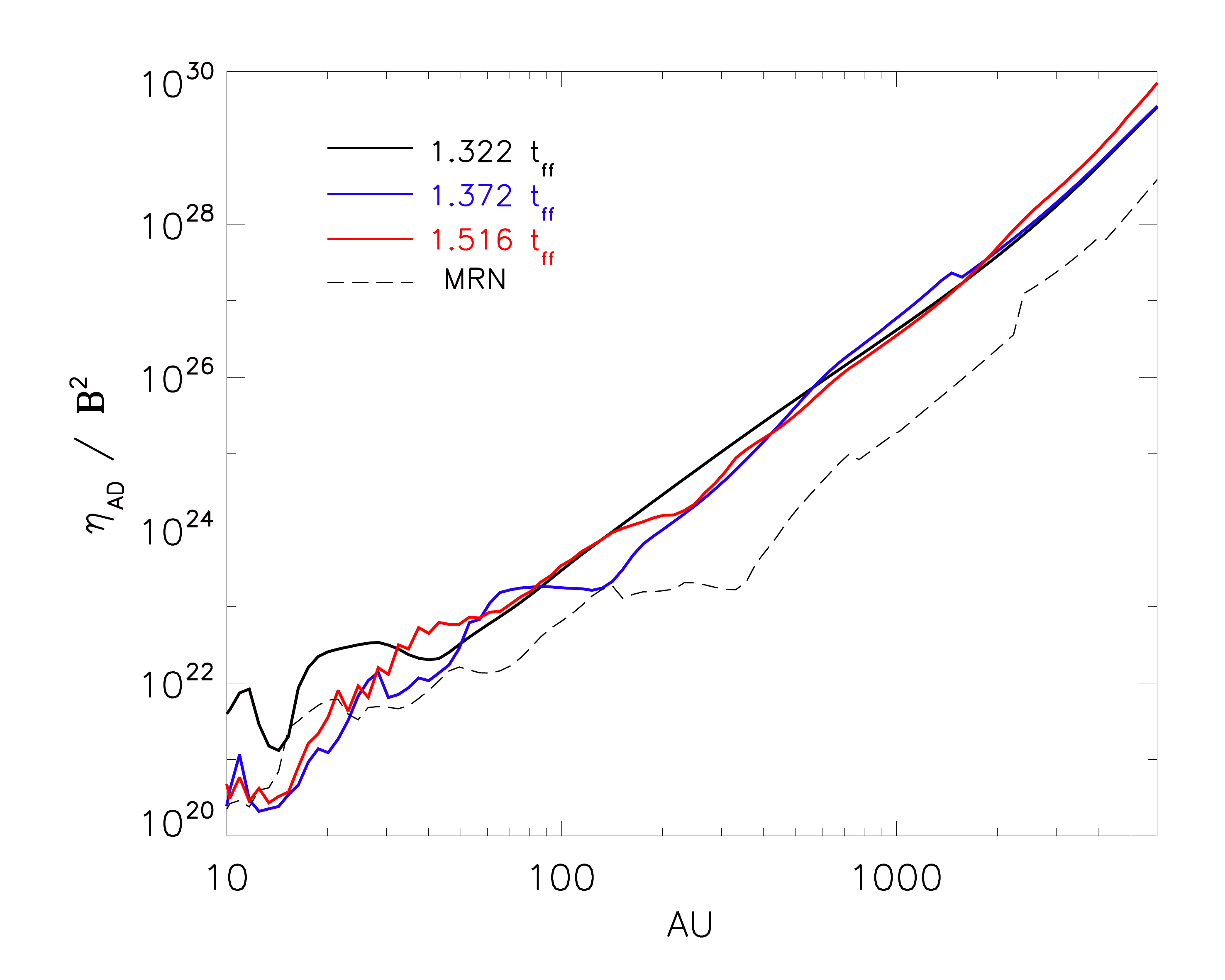}
\caption{$\eta_{\rm AD}/\bmath{B}^2$ for three different times of the 
2.4Slw-trMRN model, along with the 2.4Slw-MRN model at $1.487 t_{\rm ff}$. 
The values are plotted along the curved infall stream or the pseudo-disc.}
\label{Fig:ad_Eta}
\end{figure}

Along the equator, $B_\theta$ is essentially the vertical poloidal field; 
however, the gradient of such fields along the radial direction is much 
smaller compared to the change of $B_r$ across the $\theta=90^\circ$ 
equatorial plane, where the sign of $B_r$ is reversed at the pinching point. 
Therefore, the $r$-directional drift velocity can be further simplified 
to, 
\begin{equation}
\label{Eq:vdr_final}
{\rm v}_{{\rm d},r} \approx {\eta_{\rm AD} \over \bmath{B}^2} {B_\theta \over r} {\partial B_r \over \partial \theta}~,
\end{equation}
which depends mainly on the ambipolar coefficient 
$\eta_{\rm AD}/\bmath{B}^2$ and the tension term of the Lorenz force 
$(B_\theta / r)(\partial B_r / \partial \theta)$. 
Fig.~\ref{Fig:ad_Eta} shows that the ambipolar coefficient 
at different times in the tr-MRN case does not change much 
over a wide region ($\gtrsim$30~AU) outside the disc, which cannot be 
responsible for the expansion of the decoupling region over time. 
However, their values in the envelope are indeed a factor of 10 or 
more larger than in the MRN case, 
which explains the absence of fully decoupled regions in the MRN case 
shown in Fig.~\ref{Fig:2.4Slw-MRN_D}. 

\begin{figure*}
\includegraphics[width=1.03\textwidth]{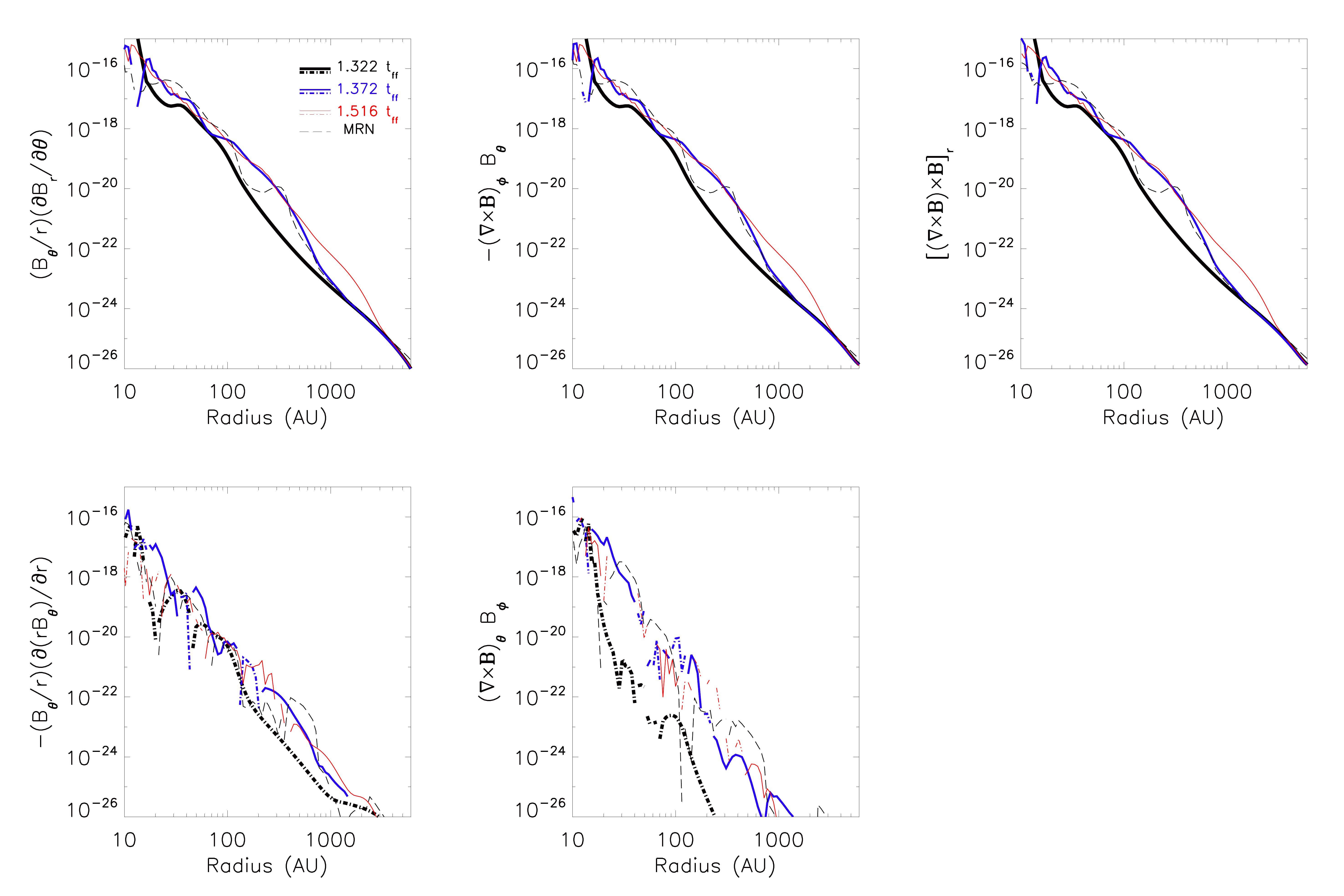}
\caption{Lorenz force terms $(B_\theta / r)(\partial B_r / \partial \theta)$,
($-\nabla \times \bmath{B}$)$_\phi B_\theta$, 
[($\nabla \times \bmath{B}$)$\times \bmath{B}$]$_r$, 
$(B_\theta / r)(\partial r B_\theta / \partial r)$, and 
($\nabla \times \bmath{B}$)$_\theta B_\phi$ for the three different times 
of the 2.4Slw-trMRN model, along with the 2.4Slw-MRN model 
at $1.487 t_{\rm ff}$. Solid lines represent positive values and 
dot-dashed ones represent negative values. 
The values are plotted along the curved infall stream or the pseudo-disc.}
\label{Fig:ad_Force}
\end{figure*}

After eliminating the role of $\eta_{\rm AD}/\bmath{B}^2$, we are 
left with only one possibility to account for the expansion of the 
decoupling region --- the tension force 
$(B_\theta / r)(\partial B_r / \partial \theta)$.
In Fig.~\ref{Fig:ad_Force}, we plot the different terms of the Lorenz force 
based on Eq.~\ref{Eq:vdr_deriv}--\ref{Eq:vdr_final}, along a line through 
the curved infall stream (or pseudo-disc, see sketch in the bottom panel of 
Fig.~\ref{Fig:ad_Vel}) that properly captures 
the positions with the most severe pinching of magnetic field lines. 
When comparing curves of different times, the tension term 
$(B_\theta / r)(\partial B_r / \partial \theta)$ clearly shows larger 
values in the corresponding decoupling regions. For instance, at 
$t \approx 1.372 t_{\rm ff}$, the tension term between 100-600~AU 
is about 10 times higher than that at an early time 
$t \approx 1.322 t_{\rm ff}$ that has a decoupling region between 
20--100~AU. Hence, the total decoupling zone at $t \approx 1.372 t_{\rm ff}$ 
is combined into 20--600~AU, which matches the velocity separation in 
Fig.~\ref{Fig:ad_Vel}. A similar reasoning applies to 
$t \approx 1.516 t_{\rm ff}$ that yields an even wider decoupling 
zone between 20--2000~AU; this again matches the above-mentioned 
velocity separation.

\begin{figure}
\includegraphics[width=1.0\columnwidth]{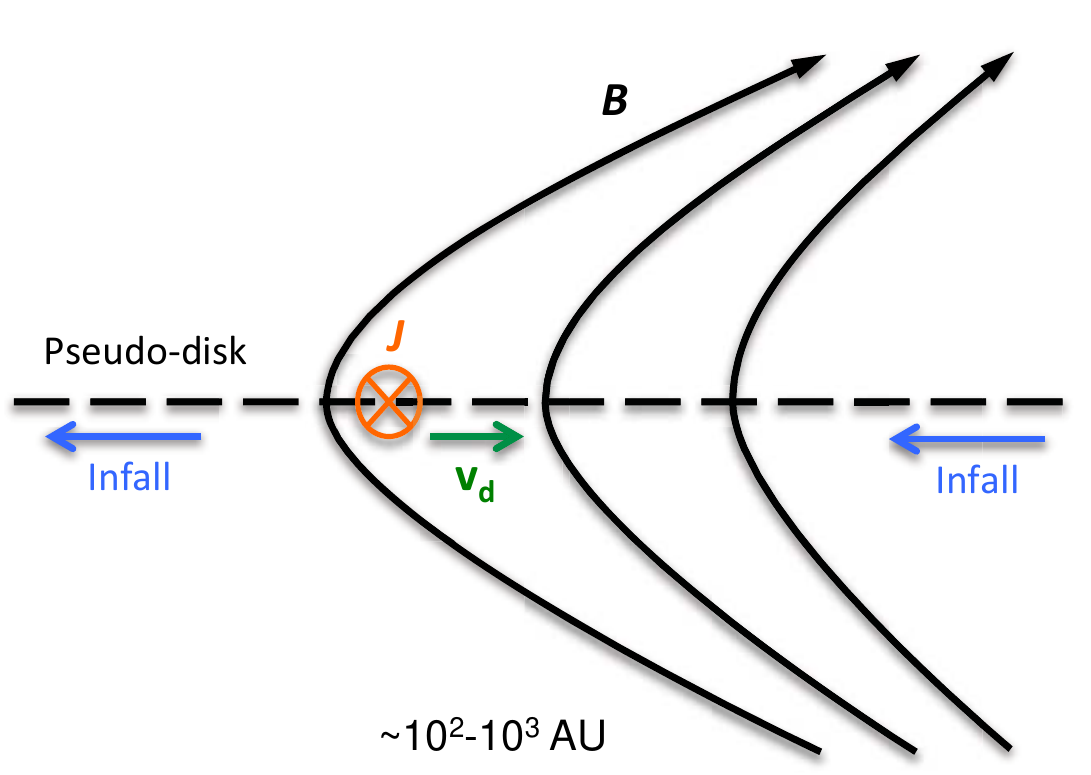}
\caption{Illustration of AD drift in the collapsing envelope, in which 
$\bmath{J}={c \over 4\pi} \nabla \times \bmath{B}$ is the current induced 
by the field pinching across the pseudo-disc, and $\bmath{\rm v_d}$ is 
the ambipolar drift velocity.}
\label{Fig:ad_Sketch}
\end{figure}

Note that the three terms $(B_\theta / r)(\partial B_r / \partial \theta)$, 
($-\nabla \times \bmath{B}$)$_\phi B_\theta$, and 
[($\nabla \times \bmath{B}$)$\times \bmath{B}$]$_r$ are almost identical 
to each other, with only very small discrepancies in the inner 20~AU. 
This validates the approximations we made in 
Eq.~\ref{Eq:vdr_deriv}--\ref{Eq:vdr_final}. 
Indeed, the gradient of $B_\theta$ and the whole term 
($\nabla \times \bmath{B}_\theta$)$B_\phi$ are more chaotic 
and orders of magnitude smaller than their respective counterparts. 
The dominance of tension force among the Lorenz force terms and 
its expansion in radius over time therefore proves 
the collapse triggered field pinching to be the main reason for 
the large expanding decoupling region in the envelope. 
The mechanism of this type of field decoupling is summarized in the 
illustration Fig.~\ref{Fig:ad_Sketch}.

\begin{figure}
\includegraphics[width=\columnwidth]{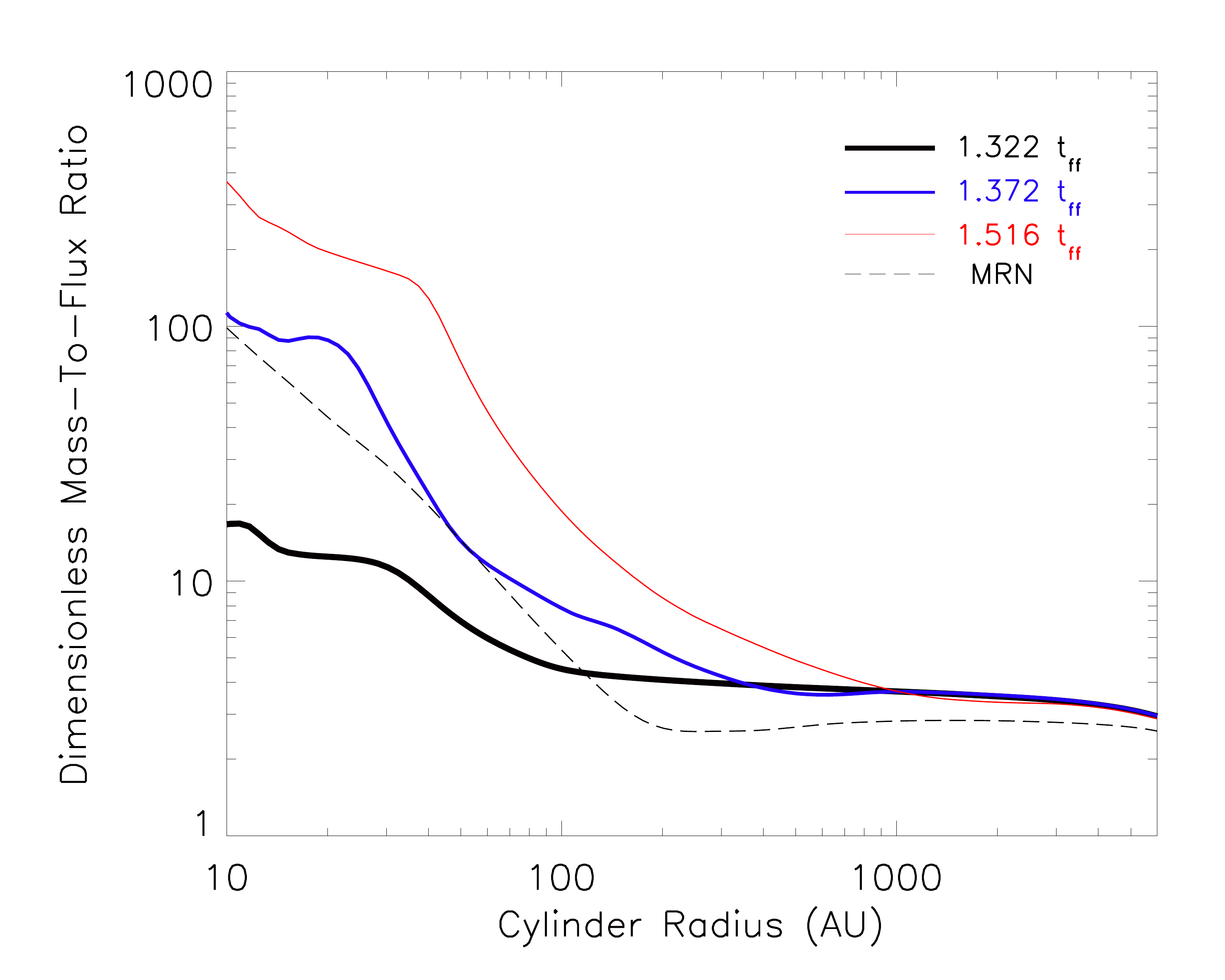}
\caption{Dimensionless mass-to-flux ratio $\lambda$, 
for three different times of the 2.4Slw-trMRN model, 
along with the 2.4Slw-MRN model at $1.487 t_{\rm ff}$.}
\label{Fig:ad_Mtf}
\end{figure}

As collapse proceeds, the efficient field decoupling prevents 
a large fraction of the magnetic flux from moving inward with 
the collapsing neutral flow. We thus expect the mass-to-flux 
ratio $\lambda$ to increase over time, which is exactly the case 
in Fig.~\ref{Fig:ad_Mtf}. $\lambda$ increases from 2--3 at the 
cloud edge to $\sim$10, $\sim$100, and $\sim$200 near the centre 
at time $1.322 t_{\rm ff}$, $1.372 t_{\rm ff}$, and $1.516 t_{\rm ff}$, 
respectively. In comparison, $\lambda$ in the MRN model at 
time $1.487 t_{\rm ff}$~kyr is a few times lower than that 
of the tr-MRN cases outside $\sim$100~AU. For regions inside $\sim$100~AU,
we compare the MRN case with the tr-MRN case at time $1.372 t_{\rm ff}$, 
in that they are similar in terms of total star plus disc mass.

\begin{figure*}
\includegraphics[width=1.03\textwidth]{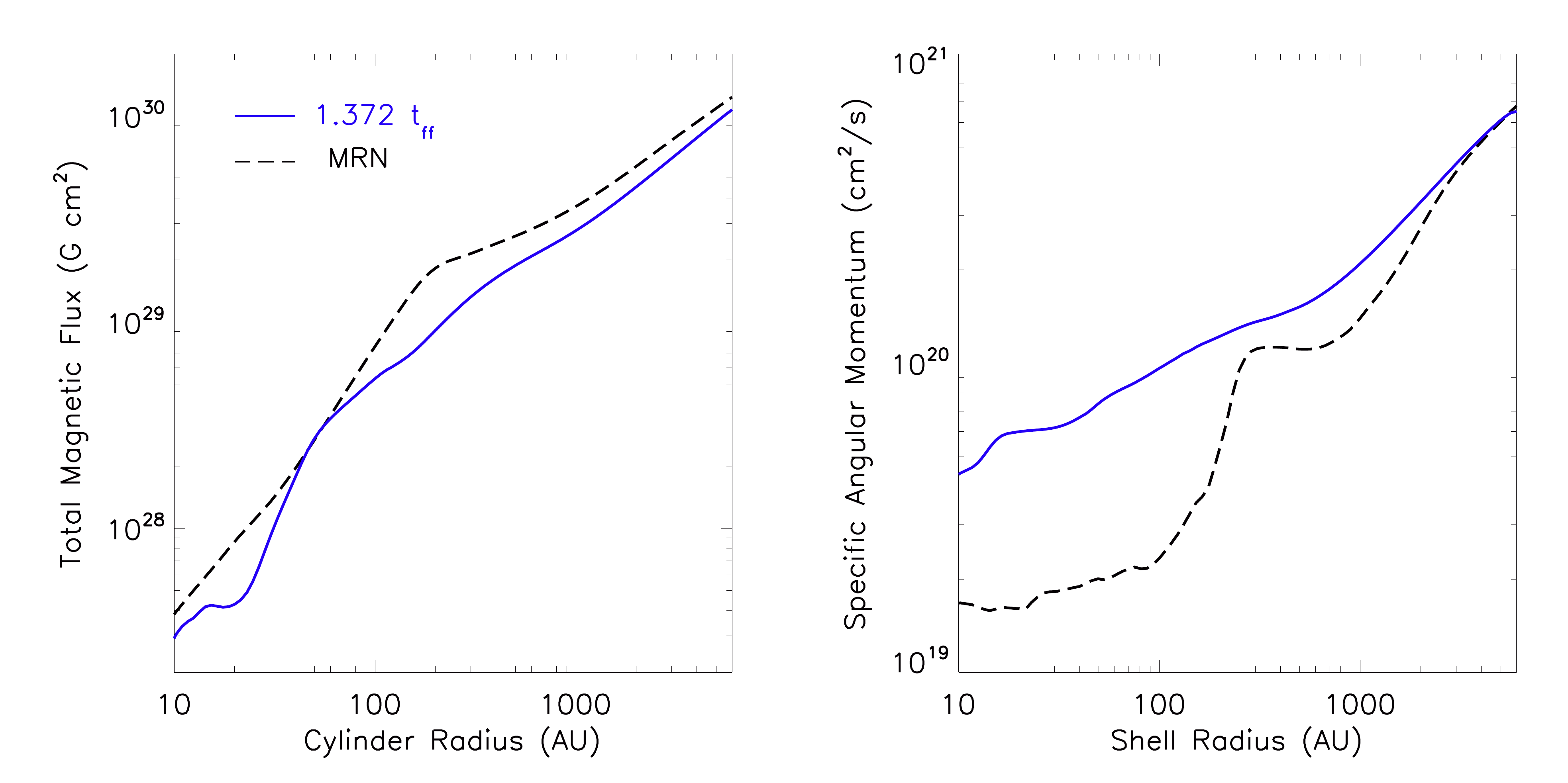}
\caption{Total magnetic flux inside cylinders of different radii 
(left), and specific angular momentum inside shells at different radii 
(right), for the 2.4Slw-trMRN model at $1.372 t_{\rm ff}$ and the 
2.4Slw-MRN model at $1.487 t_{\rm ff}$. The two chosen frames have a 
similar total mass of star plus disc.}
\label{Fig:ad_fluxAM}
\end{figure*}

The mass-to-flux ratio $\lambda$ in the tr-MRN model at $1.372 t_{\rm ff}$
is no less than the MRN model at $1.487 t_{\rm ff}$. It is indeed 
caused by a difference in magnetic flux (inside a given cylinder) 
shown in Fig.~\ref{Fig:ad_fluxAM}, 
since the two cases are comparable in mass evolution. 
Inside $\sim$2000~AU, the specific angular momentum is already 
a factor of a few higher in the tr-MRN case than in the MRN case, 
which is consistent with the result in Zhao+16. Such a difference is 
even larger in the inner $\sim$200AU where DEMS reside. 
Note that the infalling gas piles up at $\sim$200--300~AU outside the 
DEMS in the MRN case, hence resulting in a small plateau in the 
specific angular momentum outside the DEMS. 
Therefore, the excessive magnetic flux in the MRN model 
both lowers the specific angular momentum of the infalling gas in a wide 
region, and results in the magnetically dominated DEMS that obstruct the 
rotation in the innermost disc forming region. 

To conclude, it is worth noting that the decoupling region occurs 
preferentially in the vicinity of the infall channel where 
field pinching is the strongest 
\citep[similar to][]{KrasnopolskyKonigl2002}. In an axisymmetric collapse, 
such a channel is along the equator. Larger than $\sim$3$^\circ$ (with 
respect to the origin) above or below the equator, the separation of infall 
velocity between the magnetic field and the neutrals nearly disappears in 
either the MRN or tr-MRN models. 
Besides, the AD decoupling of magnetic fields triggered by collapse 
requires a lack of VSGs before the onset of discs, 
in order to provide enough ambipolar diffusivity in the envelope. 
Therefore, the process of removing VSGs must have already taken 
place in the 10$^3$~AU scale envelope or even in the 
prestellar phase.

\subsection{Trapping and Escaping of Early DEMS in Discs}
\label{S.DEMS_dynamic}

As shown in \citet{Zhao+2011} and \citet{Krasnopolsky+2012} for 
both the ideal MHD limit and non-ideal MHD cases, 
DEMS are inevitable structures once magnetic flux is by any means 
decoupled from the accreted matter (onto the star). 
In \S~\ref{S.model-MRN}, we have demonstrated the catastrophic role 
of DEMS in suppressing disc formation and obstructing disc rotation 
in the MRN cases. However, in the tr-MRN models, the DEMS are 
still present, but are less destructive and play a less important 
role.

\begin{figure*}
\includegraphics[width=1.08\textwidth,left]{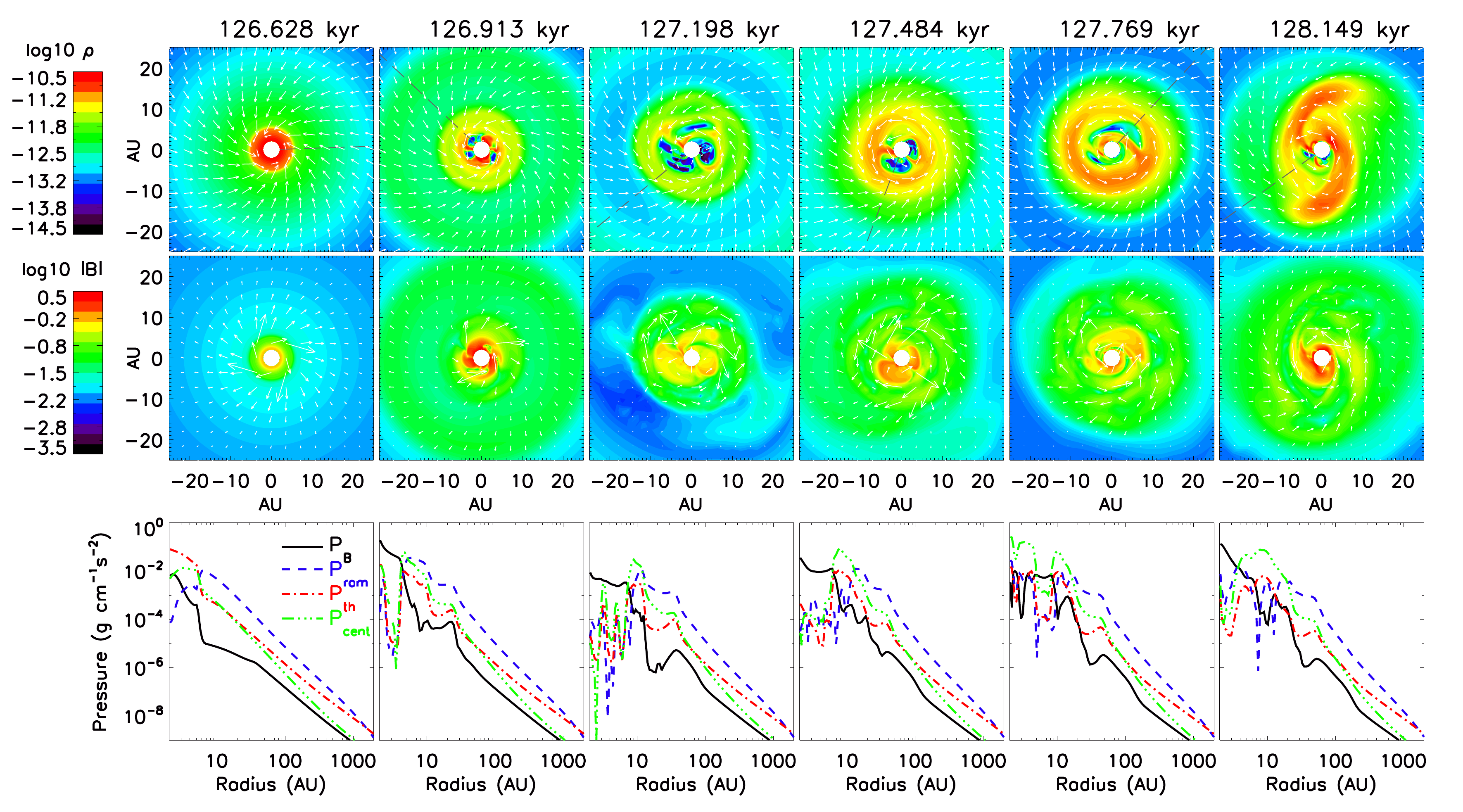}
\caption{Early evolution of disc and DEMS for 2.4Slw-trMRN model. 
Top row: distribution of mass density $\rho$ (g~cm$^{-3}$) and 
velocity field vectors within 0.336$^{\circ}$ of the equator. 
Middle row: distribution of magnetic field strength $|B|$ and magnetic 
field vectors on the same surface. 
Bottom row: thermal $P_{\rm th}$, magnetic $P_{\rm B}$, 
ram $P_{\rm ram}$, and effective centrifugal $P_{\rm cent}$ pressures 
along the same surface; the line cuts are drawn from the origin and pass 
through the low-density DEMS, with azimuthal angle $\phi$=1.9$^{\circ}$, 
137$^{\circ}$, 219$^{\circ}$, 249$^{\circ}$, 47$^{\circ}$, 
and 216$^{\circ}$, respectively (black dashed lines in the top row).}
\label{Fig:DEMS_evol}
\end{figure*}

In the tr-MRN models, the DEMS are most prominent in early phases 
shortly after the first core formation, as gas in the cloud centre 
that has the lowest specific angular momentum is quickly accreted 
by the star. As shown in Fig.~\ref{Fig:DEMS_evol}, the corresponding 
decoupled magnetic flux remains in the stellar vicinity, and is 
surrounded by the early RSD formed from the infalling gas with 
higher specific angular momentum (2nd--5th panels from the left).

At the beginning, the axisymmetric first core (1st panel from the left) 
breaks into filamentary accretion flows (2nd panel from the left) 
in between DEMS under the so-called magnetic interchange instability 
\citep{Parker1979,Kaisig+1992,StehleSpruit2001,Krasnopolsky+2012}. 
The DEMS (high magnetic field strength and low density regions 
dominated by $P_{\rm B}$) tend to expand as the central star grows 
in mass and more magnetic flux is decoupled (3rd panel from the left); 
however, the structures are quickly contained and squeezed by the 
$\sim$10~AU initial ring-shaped RSD (4th panel from the left). 
The RSD (high density disc regions dominated by $P_{\rm cent}$), which 
is unrelated to the already disrupted first core, 
is formed by assembling infalling gas with large enough centrifugal radius, 
thanks to the efficient decoupling of magnetic flux at larger radii 
that allows gas to retain enough angular momentum along their way 
to the centre (as discussed above in \S~\ref{S.fieldPinch}). 
Within a few 100~yr, the RSD quickly grows and becomes massive enough 
($\gtrsim$50\% of stellar mass) to be self-gravitating 
(Toomre $Q$ parameter $\sim$0.2, see Fig.~\ref{Fig:MasQ_evol}) and 
develops spiral structures that open up escape channels for DEMS 
(6th panel from the left). 

At later times, the DEMS are less obvious and become even less 
disruptive to disc evolution, as most magnetic flux is excluded 
from the thermally- and rotationally-dominated disc.\footnote{
Note that the RSD and the pseudo-disc in the 2.4Slw-trMRN model are 
non-coplanar (Fig.~\ref{Fig:ad_Vel}, see also \ct{Li+2014}). 
Despite the unusual geometry, 
the infall stream along the pseudo-disc still lands relatively far from 
the central star, with a radial distance between 10--15~AU (centrifugal 
radius, see also \S~\ref{S.shrinkDisc} for detailed discussions).} 
In other words, a sizeable RSD ($\gtrsim$10~AU) further prevents 
magnetic flux from reaching the very centre and minimizes the 
development of DEMS around the star. Recall that to form 
and retain such a RSD requires enough angular momentum 
in the infalling gas, which is achieved in the tr-MRN models. 
We will show a shrinking disc example in the MRN model in 
\S~\ref{S.shrinkDisc} where the initial RSD can quickly shrink 
into the central sink hole by accreting gas with low specific angular 
momentum; in such a case, large DEMS re-appear at later times. 

It is worth clarifying that the development of DEMS is a natural 
result of magnetic flux conservation and does not depend on 
the detailed decoupling mechanisms 
\citep[Ohmic, AD, or Hall; see][]{Krasnopolsky+2012}. 
The results in this section also imply that 
\begin{enumerate}
\item the first hydrostatic core \citep{Larson1969} is prone 
to disruption by magnetic instabilities and is unlikely the 
origin of the RSD formed later, in contrast to the claims by 
previous studies \citep{Bate1998,Bate2010,Bate2011,MachidaMatsumoto2011};
\item the onset of RSD takes place outside the DEMS region, which is 
enabled by the sufficient angular momentum in the infalling gas;
\item after a sizeable RSD ($\gtrsim$10--15~AU) is in place, most 
magnetic flux can be further kept outside the disc, 
or the centrifugal radius of the infalling gas (definition given in 
Eq.~\ref{Eq:Rcent}) if it is smaller than the disk radius.
\end{enumerate}

\subsection{Disc Fragmentation \& Formation of Multiple Systems}
\label{S.FragMulti}

Until recently, disc fragmentation has shown to be difficult 
if the initial core is moderately or strongly magnetized 
($\lambda \lesssim 10$) \citep[e.g.,][]{LewisBate2017}. In contrast, 
we find that fragmentation can indeed occur on spirals or rings 
in the tr-MRN models (Table~\ref{Tab:models}) with a relatively 
strong initial magnetic field ($\lambda=4.8$). 
Note that the spiral or ring structures themselves are clear signs of 
high specific angular momentum entering the inner disc-forming region. 

The general criterion for disc stability and spiral formation 
can be estimated by Toomre's Q parameter \citep{Toomre1964}, 
\begin{equation}
\label{Eq:ToomreQ}
Q={c_{\rm s} \kappa \over \pi G \Sigma}~,
\end{equation}
where $c_{\rm s}$ is the sound speed within the disc, $\kappa$ is the 
epicyclic frequency which equals to $\Omega$ (angular rotation frequency) 
for a Keplerian disc, and $\Sigma$ is the disc surface density. 
When $Q\lesssim1$, the disc becomes susceptible to the growth of 
spiral wave modes, which is exactly the case in the early phases 
of all tr-MRN models. By substituting $\Omega^2 \approx GM_*/r^3$, 
disc mass $M_{\rm d} \approx \pi r^2 \Sigma$, and disc scale height 
$H \approx c_s/\Omega$, Eq.~\ref{Eq:ToomreQ} can be further simplified 
to the following approximation,
\begin{equation}
\label{Eq:ToomreQest}
Q \approx 2{M_* \over M_{\rm d}} {H \over R_{\rm d}}~
\end{equation}
\citep{Tobin+2016}, where $M_*$ is the stellar mass and $R_{\rm d}$ 
the disc radius. To identify the disc in our simulations, we 
use the following criteria \citep[similar to][]{Joos+2012}:
\begin{enumerate}
\item Density is above certain critical value $\rho_{\rm cr}$, i.e.,
$\rho$>$\rho_{\rm cr}$=$10^{-13}$~g~cm$^{-3}$;
\item Azimuthal velocity dominates over radial velocity, i.e.,
${\rm v}_\phi$>$f_{\rm thres} {\rm v}_r$;
\item Rotational support dominates over both thermal and magnetic support, 
i.e., $\rho {\rm v}_\phi^2/2$>$f_{\rm thres} P_{\rm th}$, and 
$\rho {\rm v}_\phi^2/2$>$f_{\rm thres} P_{\rm mag}$.
\end{enumerate}
We choose $f_{\rm thres}=2$ in our analysis. Note that the criterion 
of low magnetic support is generally not important, because the high density 
regions satisfying criterion (i) generally have plasma-$\beta$ on the order 
of $\sim$10$^1$--10$^4$ (e.g., Fig.~\ref{Fig:Disc_Pres}).\footnote{
The high density part of the spiral structures generally have plasma-$\beta$ 
on the order of $\sim$10$^1$, while the plasma-$\beta$ in the disc is 
at least a few 10$^2$.} For the same reason, the magnetic 
Toomre parameter \citep{KimOstriker2001,Seifried+2013} is almost 
identical to the standard Toomre $Q$ parameter under the above criteria. 

We apply these analyses to both the non-fragmenting (2.4Fst-trMRN)
and fragmenting (4.8Slw-trMRN \& 4.8Fst-trMRN) models. 
We confirm that the criterion $Q\lesssim1$ alone is insufficient 
for the occurrence of fragmentation; the instability induced spiral structures 
also have to accrete materials from the envelope more rapidly than it could 
transport them inward \citep{Kratter+2010}. 
In our simulation, we observe that the fragments usually form 
at the local centrifugal barrier (sections of the spiral or ring 
structures, shown below) where the infall material piles up.

\subsubsection{Spiral Structures and Centrifugal Barrier: \\ 
Model 2.4Fst-trMRN}

\begin{figure*}
\includegraphics[width=1.0\textwidth]{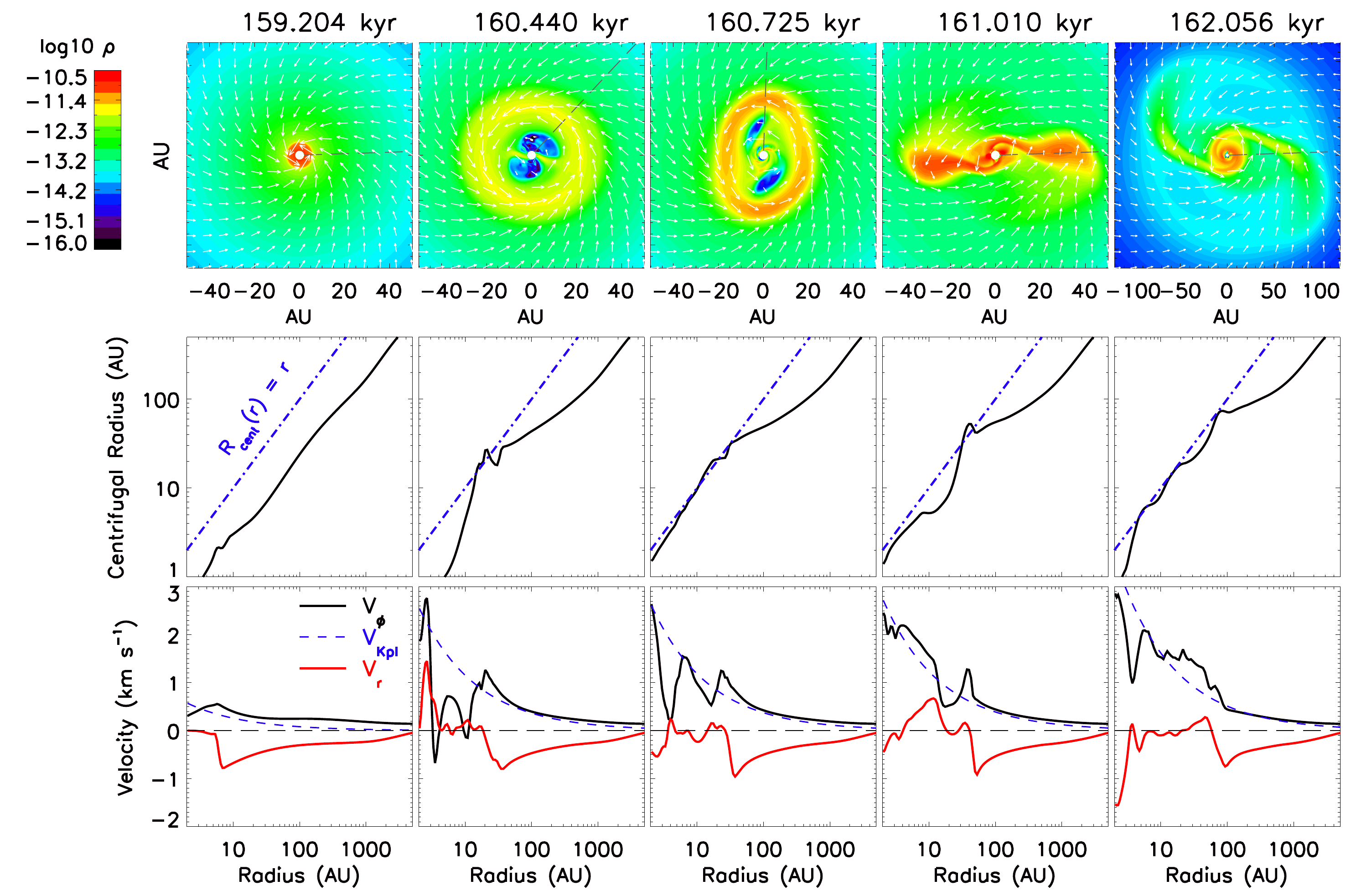}
\caption{Early evolution of spiral structures for 2.4Fst-trMRN model. 
Top row: distribution of mass density $\rho$ (g~cm$^{-3}$) 
and velocity field vectors within 0.336$^{\circ}$ of the 
equator. Middle row: estimated centrifugal radius. 
Bottom row: velocity profile along the same surface; 
the line cuts are drawn from the origin with azimuthal angle 
$\phi$=1.9$^\circ$, 47$^\circ$, 88$^\circ$, 1.9$^\circ$, and 1.9$^\circ$, 
respectively (black dashed lines in the top row).}
\label{Fig:Spiral_evol}
\end{figure*}

In the 2.4Fst-trMRN model (higher magnetization), no obvious 
fragmentation takes place throughout the simulation even though 
Toomre $Q<1$ (see detailed discussion in \S~\ref{S.fragStable}); 
instead, a grand spiral structure steadily develops in the central 
100~AU region. As shown in Fig.~\ref{Fig:Spiral_evol}, the early evolution 
is very similar to the 2.4Slw-trMRN model (slower rotation) above. 
The $\sim$5~AU first core (1st panel from the left) is quickly 
disrupted within 
$\sim$600~yr and the inner $\sim$10~AU region is occupied by the DEMS. 
Surrounding the DEMS, a RSD is assembled and gradually grows as gas with 
high specific angular momentum falls in (2nd panel from the left). 
This stage lasts about $\sim$1.2~kyr until the RSD grows more massive 
than the central star and becomes gravitationally unstable. As shown 
in Fig.~\ref{Fig:MasQ_evol}, the disc mass catches up and overtakes the 
stellar mass around $t\sim160$~kyr, and the $Q$ parameter decreases 
rapidly from $\sim$1.0 to $\sim$0.2. 
In the 3rd panel from the left, the RSD is compressed by the infall into 
a ring shape and starts to wobble. The northern and southern sections of 
the ring accumulates materials faster than other ring sections, 
which rapidly breaks the ring symmetry. Within 1/4 orbit, 
the two mass-gaining regions are turned into two spirals 
while the other ring sections fall to the centre. During the process, 
the DEMS find open channels to escape from the inner region, similar to 
the 2.4Slw-MRN case. Thereafter, accretion streams are able to wrap around 
the central star to form a well-defined disc. In another $\sim$1000~yr, 
the whole disc-spiral structure extends to $\sim$100~AU. 

The development of these rotationally supported structures 
(disc, ring and spiral) is essentially controlled by the 
centrifugal radii (angular momentum) of infalling gas parcels. 
Approximately, the expected centrifugal radius $R_{\rm cent}(r)$ for 
gas in any given shell $r$ can be expressed as,
\begin{equation}
\label{Eq:Rcent}
R_{\rm cent}(r)={j(r)^2 \over G (M_*+\int_{\rm r'<r} M(r'))}~,
\end{equation}
(Zhao+16), where $M_*$ is the stellar mass and $j(r)$ is the averaged 
specific angular momentum in that shell; however, when computing $j(r)$,
we only consider materials within $\pm$45$^\circ$ above and below the 
equator (45$^\circ$<$\theta$<135$^\circ$) to avoid the bipolar 
outflow regions. The mass integration is still carried out using the 
total mass inside radius $r$ for an approximated spherical potential. 

The distribution of $R_{\rm cent}(r)$ is plotted in the second row of 
Fig.~\ref{Fig:Spiral_evol} for the corresponding snapshots in the top row.
Except for the first core phase ($M_*\lesssim10^{-3}~M_{\sun}$), 
the rest has a common ``plateau'' feature in the $R_{\rm cent}$ 
curve. It is a typical signature of the ``so-called'' centrifugal barrier 
(Zhao+16) because the infalling gas at different radii along the 
``plateau'' tends to arrive at the same radius $R_{\rm cent}$. 
Such centrifugal radii are around 20--30~AU (2nd panel from the left), 
30--40~AU (3rd panel from the left), 40--50~AU (4th panel from the left), 
and 70--80~AU (5th panel from the left) 
for the respective snapshots. As collapse continues to bring the envelope 
gas with higher angular momentum (initial solid-body rotation), the 
location of the centrifugal barrier moves outward. 

The centrifugal barrier can also be identified from the kinematic 
profiles. As the envelope gas falls towards the central 
region, its rotation speed ${\rm v}_\phi$ suddenly rises and infall 
speed ${\rm v_r}$ almost vanishes across the centrifugal barrier. 
Interior to the centrifugal barrier (post shock), ${\rm v}_\phi$ 
tends to decrease to the local Keplerian speed, unless the inner 
regions are dominated by the DEMS. 
It is worth pointing out that the centrifugal barrier at later times 
(4th and 5th panels from the left) coincides with the outer arm of the 
spiral structures, as compared to the rigid rings formed by piling up infall 
material shown in Zhao+16 (2D). In this sense, the centrifugal 
barrier in 3D is much less stiff or obstructive to the infall flows 
than in 2D. 

\subsubsection{Fragmentation of Spirals \& Accretion Bursts: \\
Model 4.8Slw-trMRN}

In the 4.8Slw-trMRN model with a low magnetization and slow rotation, 
fragmentation frequently occurs on the outer arm of the spiral structures, 
producing transient companion clumps with tens of Jupiter mass. 
In most cases, the companion clumps spiral inward and tidally interact with 
the circumstellar disc, leading to episodic accretion events 
\citep[Fig.~\ref{Fig:Mdot_evol}; see also][]{VorobyovBasu2006,VorobyovBasu2015}.
However, the orbital evolution of companion objects is not well-followed 
in this study due to a lack of sink particle treatment (except 
for the stationary sink hole in the centre, which can violate 
the linear momentum conservation).

\begin{figure*}
\includegraphics[width=1.08\textwidth,left]{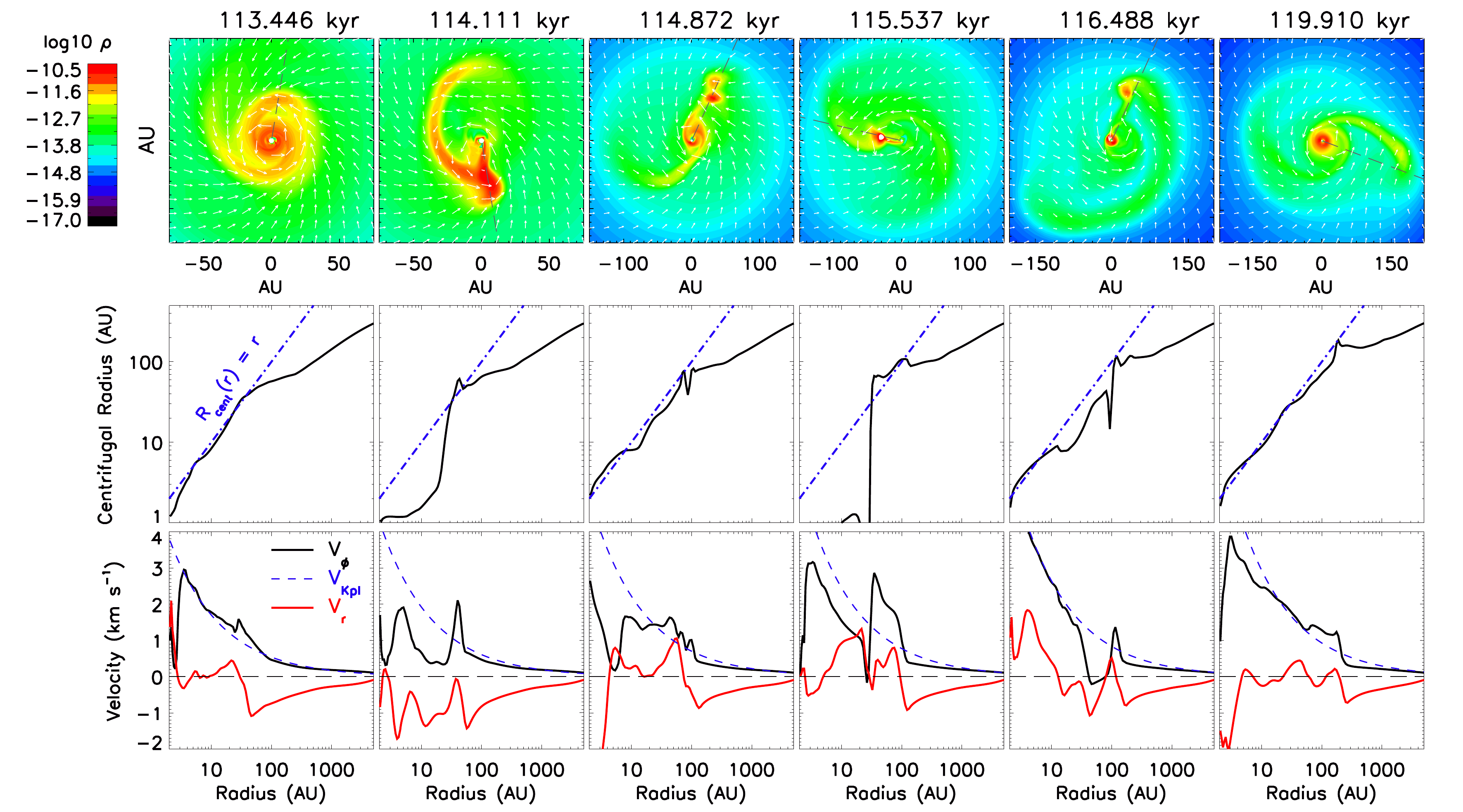}
\caption{Evolution of spiral structures for 4.8Slw-trMRN model. 
Top row: distribution of mass density $\rho$ (g~cm$^{-3}$) and 
velocity field vectors within 0.336$^{\circ}$ of the 
equator. Middle row: estimated centrifugal radius. 
Bottom row: velocity profile along the same surface; 
the line cuts are drawn from the origin with azimuthal angle 
$\phi$=81$^\circ$, 279$^\circ$, 66$^\circ$, 167$^\circ$, 66$^\circ$, 
and 339$^\circ$, respectively (black dashed lines in the top row).}
\label{Fig:Spiral_frag}
\end{figure*}

The early evolution of 4.8Slw-trMRN model is similar to that of the 
2.4Slw-trMRN and 2.4Fst-trMRN models, in which the first core is 
disrupted by DEMS and an unstable RSD ($Q\sim0.3$ at 111.7--120.1~kyr, 
see Fig.~\ref{Fig:MasQ_evol}) 
forms outside the DEMS from the infalling rotating materials. 
In Fig.~\ref{Fig:Spiral_frag}, we present the phase when the 
spiral structure is already developed from the wobbling massive RSD 
(1st panel from the left, $\sim$2.8~kyr after the formation of the 
first core). 
A slight asymmetry causes the the northern spiral arm to accumulate 
materials faster than the southern arm. Such a perturbation runs away 
rapidly within half orbital time, causing the northern arm to grow 
into a massive clump as it rotates to the southern position (2nd panel 
from the left). The massive clump, about 60\% of the stellar mass 
of 0.042~M$_{\sun}$, tidally disrupts the circumstellar disc. 
It is then followed by an accretion 
burst ($\dot{M}_* \approx$ 6--7 $\times$10$^{-5}$~M$_{\sun}$/yr) 
when materials flow along the stream from the clump onto 
the central star. Part of the clump wraps around the star to form a new 
circumstellar disc, and the rest continues to rotate and stretch into 
an arc shape. After another half orbit (3rd panel from the left), 
infalling materials pile up at 
the stretched northern arc at $r$$\sim$70--80~AU, i.e., the centrifugal 
barrier, where two companion objects form with a total mass of 
$\sim$0.017~M$_{\sun}$ ($\sim$30\% of the stellar mass). The two 
companion clumps quickly merge into one clump with a mass of 
$\sim$0.022~M$_{\sun}$, which spirals inward towards the star. 
After another 1/4 orbit (4th panel from the left), the circumstellar 
disc is disrupted again. The 0.028~M$_{\sun}$ clump near the 
0.068~M$_{\sun}$ star causes another episode of accretion burst 
($\dot{M}_* \approx$ 6--7 $\times$10$^{-5}$~M$_{\sun}$/yr). 
The situation afterwards is very similar to that of the first burst, 
where part of the clump wraps into a new circumstellar disc and the 
rest stretches out into long arc-shaped streams. Another 3/4 orbit later, 
a new clump of $\sim$0.011~M$_{\sun}$ ($\sim$15\% of the stellar mass) 
forms near the $\sim$100~AU centrifugal barrier. 

The subsequent evolution repeats the previous episodes in a similar 
fashion; and after two more accretion burst 
episodes, the system evolves into a more standard disc plus spiral 
configuration which extends to $\sim$150~AU. However, the outer spiral 
arm located near the centrifugal barrier is still susceptible to the 
formation of 10$^{-2}$~M$_{\sun}$ companion clumps. 

It is worth verifying the location of the centrifugal barrier 
from both the centrifugal radius and the velocity profile 
(plotted along a line passing through the clumps). 
Basically the locations of the ``plateau'' on the 
$R_{\rm cent}$ curve and the sharp change in ${\rm v}_\phi$ and 
${\rm v_r}$ coincide well with each other; and the location of 
such a centrifugal barrier also coincides with that of the companion 
clumps in the top panels. This indicates that the companion clumps 
are indeed formed from piling up infall material at 
the centrifugal barrier. 
Note that the curves of centrifugal radius inner to the centrifugal 
barrier are heavily affected by the companion objects at later times. 

\subsubsection{Fragmentation of Rings \& Multiple Systems: \\
Model 4.8Fst-trMRN}

\begin{figure*}
\includegraphics[width=1.08\textwidth,left]{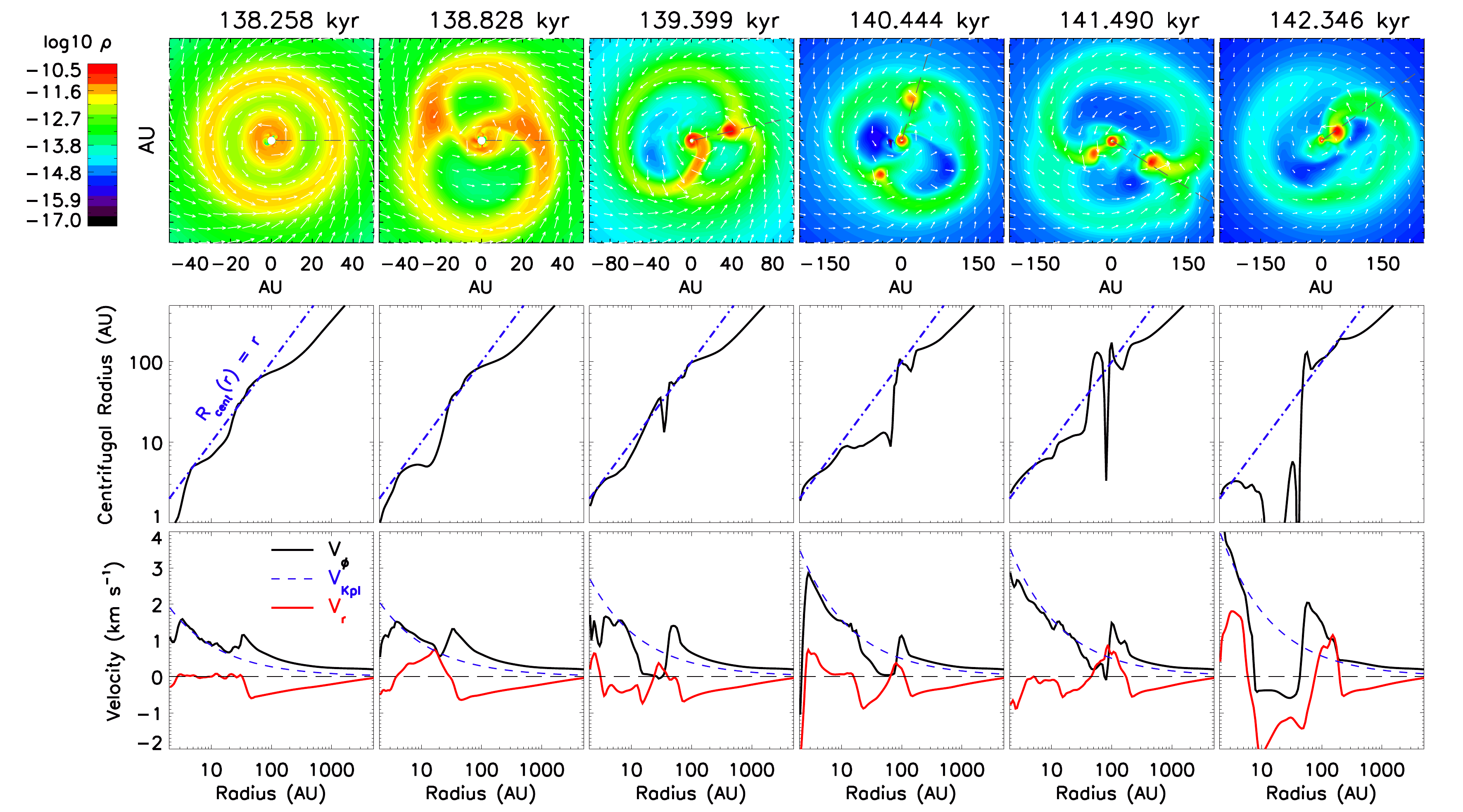}
\caption{Evolution of spiral and ring structures for model 4.8Fst-trMRN.
Top row: distribution of mass density $\rho$ (g~cm$^{-3}$) and velocity 
field vectors within 0.336$^{\circ}$ of the equator. 
Middle row: estimated centrifugal radius. 
Bottom row: velocity profile along the same surface; 
the line cuts are drawn from the origin with azimuthal angle 
$\phi$=1.9$^\circ$, 1.9$^\circ$, 13$^\circ$, 73$^\circ$, 328$^\circ$, 
and 36$^\circ$, respectively (black dashed lines in the top row).}
\label{Fig:Ring_frag}
\end{figure*}

In the faster rotating model 4.8Fst-trMRN, fragmentation of rings 
(formed due to high angular momentum influx) frequently produces 
multiple companion clumps. The early evolution of model 4.8Fst-trMRN 
however, has much smaller DEMS than the other cases, due to 
a lack of accretion onto the central star. Until $t=138.828$~kyr 
(2nd panel from the left of Fig.~\ref{Fig:Ring_frag}, $\sim$2.2~kyr 
after the formation of the first core), the stellar accretion rate is only 
around 4$\times$10$^{-7}$~M$_{\sun}$/yr. It is caused by the 
large angular momentum in the innermost $\sim$5~AU that 
prevents the gas from falling further below their large 
centrifugal orbits. Nevertheless, small DEMS can still develop in 
the inner 5~AU (1st panel from the left), which have disrupted the 
first core (similar to \S.~\ref{S.DEMS_dynamic}). Surrounding 
the DEMS is a $\sim$10--15~AU RSD with $\sim$0.006~M$_{\sun}$ 
formed from the gas with high specific angular momentum; 
Further out between 25--35~AU, a massive ring (0.024~M$_{\sun}$)
is growing outside the RSD by assembling gas from larger 
radii with even higher specific angular momentum. The ring has 4 times 
the mass of the inner RSD and 3 times the mass of the star, thus 
creating a gap between 15--25~AU by gradually attracting materials 
in the gap onto it (1st panel from the left). The massive ring quickly 
becomes gravitationally unstable, and wobbles and deforms within a few 
hundred years (about one orbital time). Tidal streams then develop, 
which connect the ring to the inner RSD (2nd panel from the left). 

The eastern and western sections of the ring preferentially accrete 
the infall matter. After another 1/2 orbit, the original western ring 
section stretches into an arm around the inner disc, while the original 
eastern section forms a 0.016~M$_{\sun}$ companion clump that equals 
the stellar mass (3rd panel from the left). Gas near the clump also 
tends to orbit around it. However, the companion clump fails to 
survive the close approach 1/2 orbit later; it is 
tidally disrupted by the central star and produces an accretion burst 
with $\dot{M}_*\sim$3$\times$10$^{-5}$~M$_{\sun}$/yr. Note that the 
original 40~AU ring structure keeps expanding to 100--200~AU scales 
concurrently. About 500~yr after the accretion burst 
(4th panel from the left), 
two well-separated companion clumps develop within the ring structure, 
with mass $\sim$0.008~M$_{\sun}$ (northern) and 0.010~M$_{\sun}$ 
(southern), respectively. They sum up to 2/3 of the stellar mass. 
Subsequently, the triple system undergoes dynamic interactions, 
with the northern lighter companion spiralling inward and the southern 
companion growing more massive. In 1/4 orbit (5th panel from the left), 
the masses of the triple system are 0.009~M$_{\sun}$ (eastern), 
0.022~M$_{\sun}$ (western), and 0.029~M$_{\sun}$ (primary). 
The western massive companion continues to accrete and becomes 
equal mass with the central primary (0.030~M$_{\sun}$). The lighter 
companion is tidally stretched when it rotates in between the two massive 
objects; it is eventually merged onto the western companion to 
form a 0.036~M$_{\sun}$ secondary, the same mass as the primary star
(6th panel from the left). The secondary later merges with the primary 
and leads to another accretion burst; however, since we do not have 
full sink particle treatment (except for the stationary sink hole in 
the centre), the result could otherwise be a tight binary system with 
mass ratio $\sim$1 or a hierarchical triple system \citep{Reipurth+2014}.

Throughout the evolution, the location of centrifugal barrier moves 
outward with time (middle row of Fig.~\ref{Fig:Ring_frag}), which 
matches with the radius of the large ring structure well. 
A second ``plateau'' also appears at 10~AU scale, corresponding to 
the centrifugally supported inner disc around the primary. 
Similar to the 4.8Slw-trMRN model above, the curves of centrifugal 
radius interior to the barrier are affected by the companion objects.
The velocity profiles are plotted along lines passing through 
the companion objects, which show the usual ``bump'' feature 
in ${\rm v}_\phi$ and sudden decrease in ${\rm v_r}$ for 
a typical centrifugal barrier. Again, it confirms that the 
most probable sites for the formation of companion objects are 
the centrifugal barriers where the infall material piles up. 

\subsubsection{Disc Stability Analysis}
\label{S.fragStable}

\begin{figure*}
\includegraphics[width=1.0\textwidth,left]{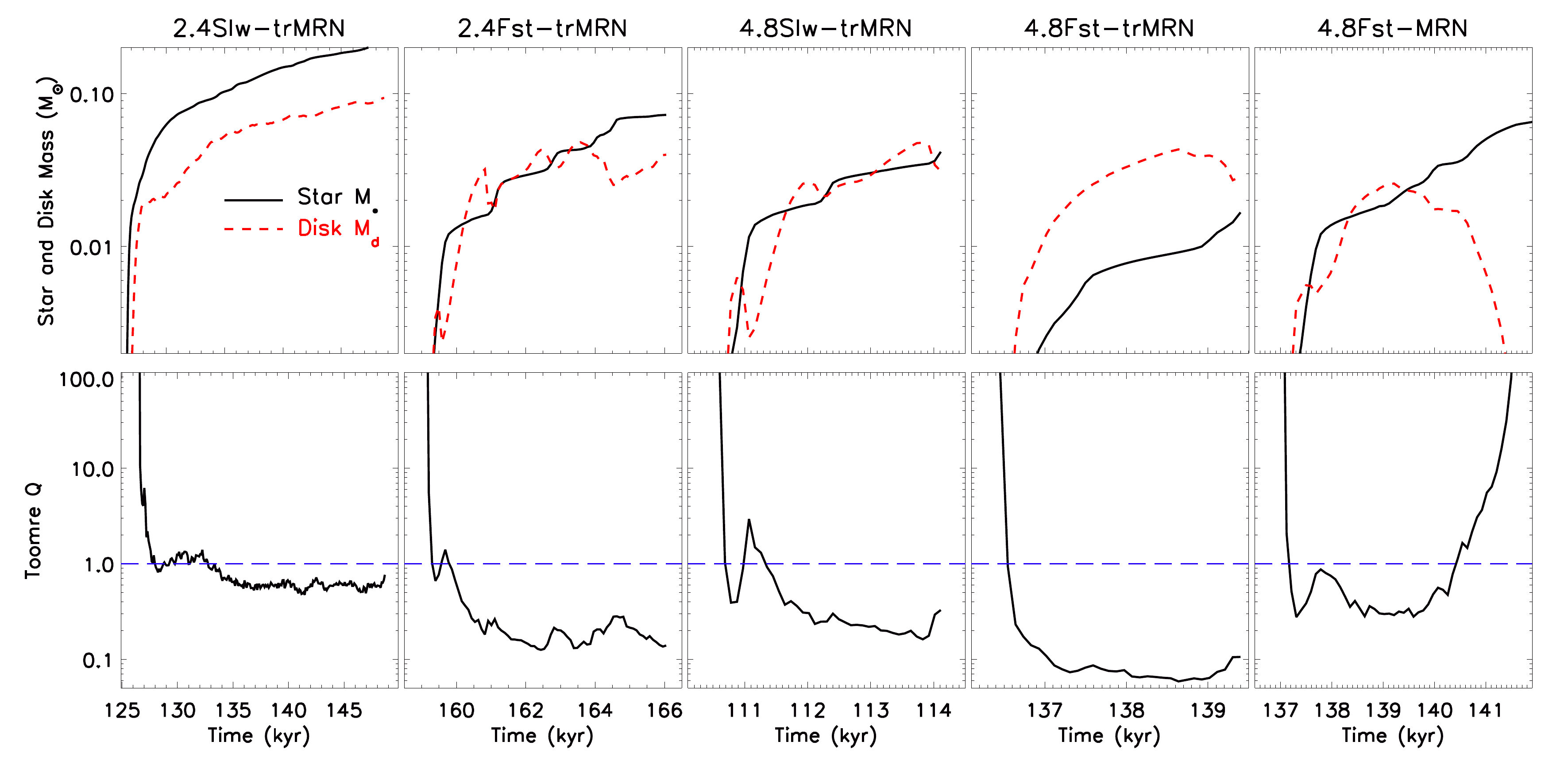}
\caption{Evolution of mass of star and disc (top row), and 
Toomre $Q$ (bottom row) for models with prominent discs. 
Dashed blue lines represent $Q=1$. 
For models with formation of companion objects, 4.8Slw-trMRN 
and 4.8Fst-trMRN, we only show the quantities before the formation 
of companions.}
\label{Fig:MasQ_evol}
\end{figure*}

\begin{figure*}
\includegraphics[width=1.0\textwidth,left]{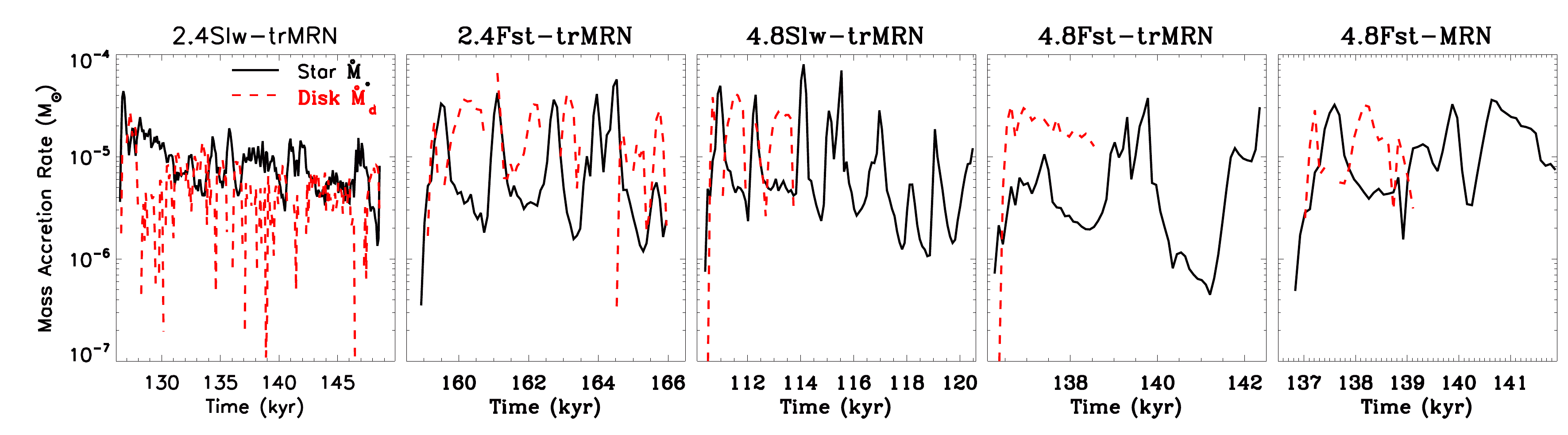}
\caption{Evolution of mass accretion rate of the star and disc 
for models with prominent discs. In comparison to Fig.~\ref{Fig:MasQ_evol}, 
we also show $\dot{M}_*$ at later times for model 4.8Slw-trMRN and 
4.8Fst-trMRN after the formation of companion objects.}
\label{Fig:Mdot_evol}
\end{figure*}

We conclude the section with a general analysis of disc stability. 
Recall that the criterion $Q<1$ only implies the growth of spiral 
waves in the disc. Indeed in Fig.~\ref{Fig:MasQ_evol}, all four 
tr-MRN models satisfy the $Q<1$ criterion, yet only the two models 
with lower magnetization ($\lambda=4.8$) show prominent fragmentation 
because the disc (or ring) is more massive and hence more unstable 
gravitationally. In these two models, new fragments often appear 
in the outer part of the spiral or ring structures, where infall 
material from the envelope piles up. In this sense, these piling-up 
spots are indeed the centrifugal barriers.

Interestingly, model 2.4Fst-trMRN and 4.8Slw-trMRN have very 
similar Toomre Q values at the early times, yet only the latter 
shows fragmentation. This is in fact consistent with the principles 
derived in \citet{Kratter+2010} (see their Eq.~1--2 and Fig.~2). 
We compare the model 2.4Fst-trMRN at $t=162.1$~kyr 
(Fig.~\ref{Fig:Spiral_evol}) and model 4.8Slw-trMRN at $t=113.4$~kyr
(Fig.~\ref{Fig:Spiral_frag}), both having a stellar mass 
of $\sim$0.03~M$_{\sun}$. While both discs have similar accretion rate 
($\sim$10$^{-5}$~M$_{\sun}$/yr) and hence similar thermal parameter 
$\zeta$,\footnote{\label{foot:zG}$\zeta$ and $\Gamma$ are 
dimensionless parameters used in \citet{Kratter+2010}.} 
the rotation speed of the spiral structure (not the envelope) 
in the 4.8Slw-trMRN model (tighter spiral) is twice faster than in 
the 2.4Fst-trMRN model (grand spiral). Accordingly, the 4.8Slw-trMRN 
model has a smaller rotational parameter 
$\Gamma$,\textsuperscript{\ref{foot:zG}} which makes 
fragmentation easier according to Fig.~2 of \citet{Kratter+2010}.
In other words, faster rotation (of the disc-spiral structure) 
promotes fragmentation.

\subsection{Failed Disc: Shrinking Disc}
\label{S.shrinkDisc}

In Table.~\ref{Tab:models}, the 4.8Fst-MRN model is of particular interest, 
in which a 10--20~AU RSD forms initially but shrinks over time and disappears 
within $\sim$3.7~kyr. The same case has been discussed in our 2D study 
Zhao+16 (see their Section 5.4). Despite the advantage of weaker 
magnetization and faster rotation, the AD in the envelope is much less 
efficient than the tr-MRN models. As a result, an increasing amount of 
magnetic flux is being dragged into the inner disc forming region as 
collapse continues. The magnetic braking in this case still operates 
efficiently to torque down the gas rotation, leading to an insufficient 
supply of specific angular momentum to maintain the current disc size. 
Therefore, the RSD shrinks in size and the disc material gradually 
falls into the central star due to a lack of rotational support. 

\begin{figure*}
\includegraphics[width=1.08\textwidth,left]{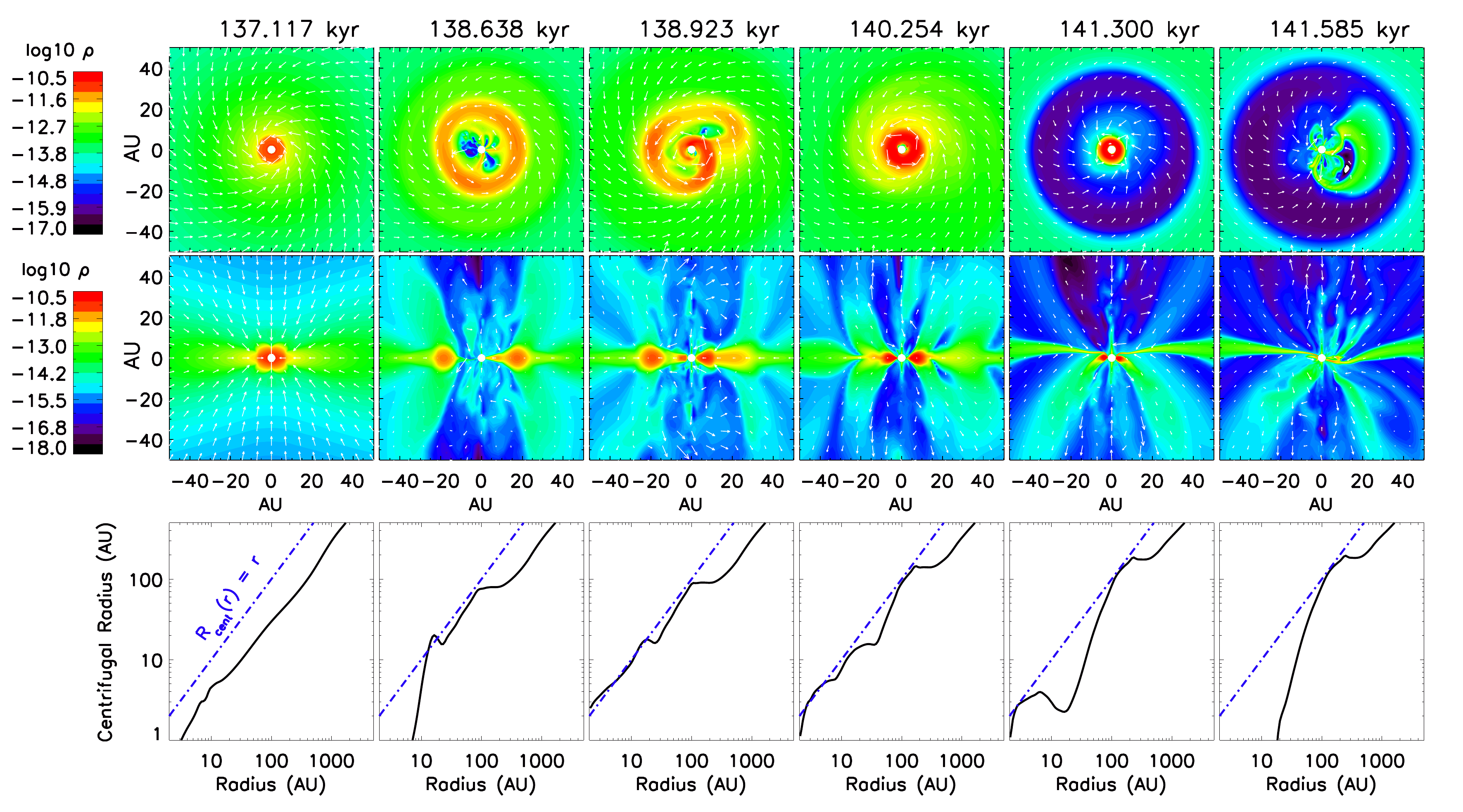}
\caption{Evolution of the shrinking disc for model 4.8Fst-MRN.
Top row: distribution of mass density $\rho$ (g~cm$^{-3}$) and 
velocity field vectors within 0.336$^{\circ}$ of the 
equator. Middle row: distribution of mass density $\rho$ (g~cm$^{-3}$) 
and velocity field vectors on the meridian plane with azimuthal angle 
$\phi$=1.9$^{\circ}$. Bottom row: estimated centrifugal radius.}
\label{Fig:Shrink_evol}
\end{figure*}

As shown in Fig.~\ref{Fig:Shrink_evol}, the early evolution of 
the 4.8Fst-MRN model is similar to the tr-MRN models: DEMS ``hatches''
in the inner $\sim$10~AU after disrupting the first core; 
and a RSD assembles between 10--20~AU (2nd panel from the left). 
The unstable ring-shaped 
RSD then deforms and wraps around the star to form the usual disc-spiral 
configuration. However, in contrast to the large spirals in the tr-MRN 
models, the spiral structure in this case quickly disappears within 
$\sim$600~yr, leaving only a compact disc of 10--15~AU radius 
(4th panel from the left). 
The suppression of spiral structures is a sign of reduced angular 
momentum influx. Notably, in the next $\sim$500~yr, the infall flow 
gradually moves away from the equatorial plane, making the pseudo-disc
to curve towards the upper hemisphere (recall that a similar phenomenon 
appears in the 2.4Slw-trMRN model as well). It is a natural outcome 
of the low angular momentum in the infalling gas; the gas flow finds 
alternative channels to reach their centrifugal radius that is much 
smaller than the disc radius. In this fashion, the disc accretes 
mass but little angular momentum, which causes the disc to shrink 
in size over time.

The change of disc radius can also be identified from the evolution 
of centrifugal radius, particularly the secondary ``plateau'' for the 
inner tens of AU radius. Across different times, the value of 
$R_{\rm cent}$ indicated by the secondary ``plateau'' well follows 
the disc radius. It first increases from $\sim$5~AU to $\sim$20~AU at 
early times (1st--3rd panels from the left), but later decreases 
from $\sim$10~AU (4th panel from the left) to $\sim$3~AU 
(5th panel from the left) and quickly vanishes in another 200~yr 
(6th panel from the left).
The lack of the inner ``plateau'' denotes the absence of RSD, which is 
exactly the case in the top row. At this time, the central region is 
instead occupied by the familiar DEMS.

\begin{figure*}
\includegraphics[width=1.0\textwidth,left]{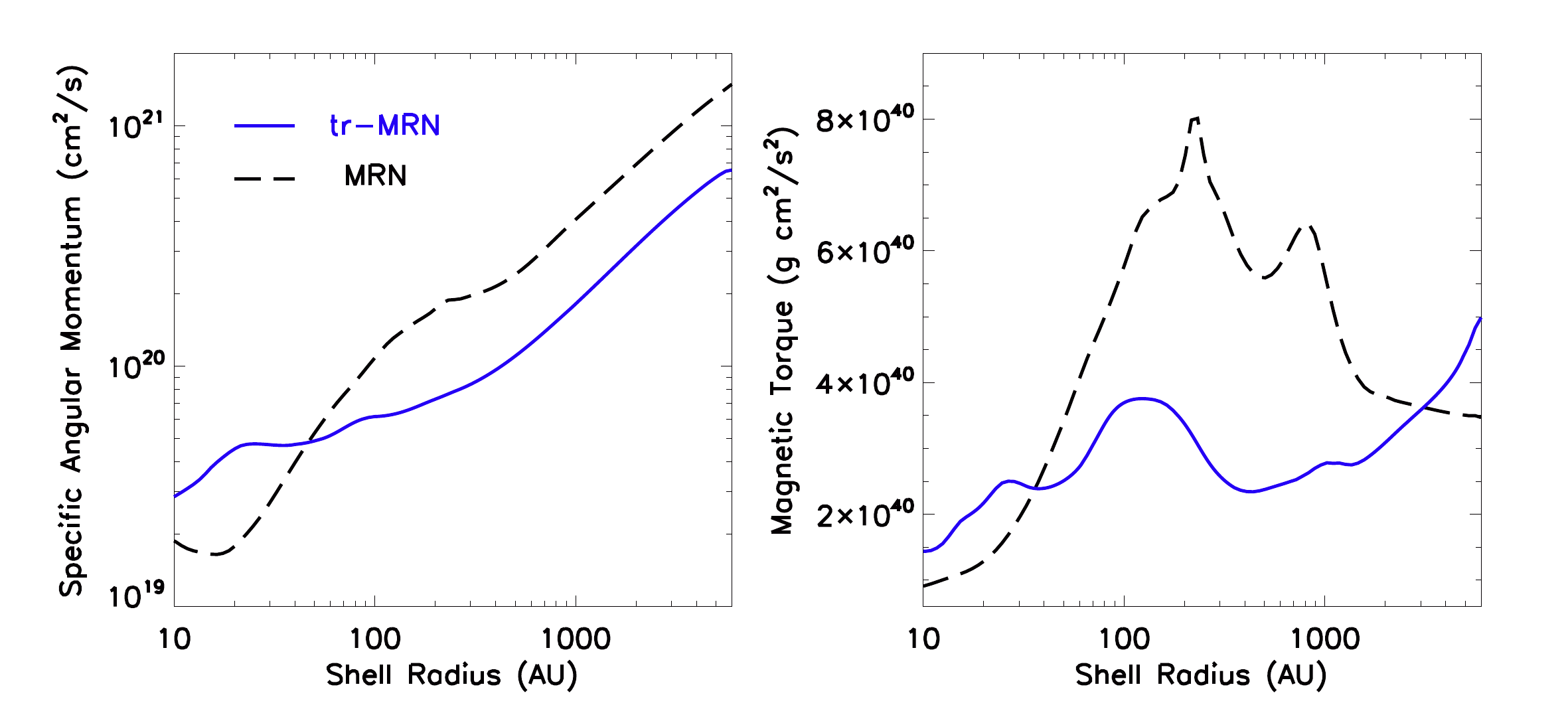}
\caption{Comparison of specific angular momentum and magnetic torque for 
4.8Fst-MRN and 2.4Slw-trMRN models at a similar evolutionary epoch 
141.3~kyr and 128.5~kyr, respectively. The two chosen frames have 
similar total mass of star plus disc ($\sim$0.06~M$\sun$).}
\label{Fig:Torq_comp}
\end{figure*}

In this 4.8Fst-MRN model, the initial advantage in rotation and 
magnetization provides sufficient angular momentum for the disc 
growth at early times; however, it is gradually taken over by the 
disadvantage in field decoupling in the envelope which determines 
the strength of magnetic braking in the disc forming region. 
We hereby compare this shrinking-disc model --- 4.8Fst-MRN with 
the steady RSD model --- 2.4Slw-trMRN (\S~\ref{S.model-trMRN}). 
Recall that the latter model has a stronger magnetization and a 
slower rotation, but no VSGs. At a similar evolutionary stage 
for the two models (similar total mass of star plus disc), 
Fig.~\ref{Fig:Torq_comp} shows the specific angular momentum 
and magnetic torque for spherical shells at different radii.
Here, we consider only the magnetic tension term while 
computing the magnetic torque:
\begin{equation}
\mathcal{N}_t(S)={1 \over 4\pi} \int_S (\bmath{r} \times \bmath{B})(\bmath{B} \cdot {\rm d}\bmath{S})~
\end{equation}
(Zhao+16), where $\bmath{S}$ is the shell surface at radius $r$. 
The contribution of the magnetic pressure term is generally 
much smaller \citep{MatsumotoTomisaka2004}. 

The magnetic torque in the 4.8Fst-MRN case peaks between 200--300~AU, 
near which the specific angular momentum curve flattens 
\citep[conservation of angular momentum;][]{Goodman+1993,Belloche2013}.
It implies that the magnetic torque is enhanced by the increasing 
rotational motion there. The magnitude of magnetic torque in the 
4.8Fst-MRN case is also a few times higher than the 2.4Slw-trMRN case, 
indicating a stronger magnetic braking in the former. Consequently, 
the specific angular momentum inside $\sim$200~AU decreases more 
rapidly in the MRN model. Such a quick slowing down in gas rotation 
in turn weakens the magnetic braking. In contrast, the strength of 
magnetic braking keeps at a relatively low level throughout the core. 
As a result, the decrease in specific angular momentum is more gradual.
Even in the inner tens of AU, there is still an adequate amount of 
angular momentum to maintain the RSD. Therefore, the effect of removing 
VSGs is more significant on maintaining a long-lived RSD than that of 
increasing the initial core rotation or mass-to-flux ratio 
(for $\lambda$ of a few).

\subsection{Outflow}
\label{S.outflow}

The formation and evolution of RSDs are naturally accompanied by 
magnetically-driven outflows. However, most outflows in our 3D models 
have multiple components, some of which can be asymmetric. 
Thus, different launching mechanisms may be in play for different 
outflow components. Note that we use an Alfv{\'e}n d$t$ 
floor\footnote{The minimum density allowed for any given cell is set to 
${|\bmath{B}|^2 \over 4\pi (|\Delta x|_{\rm min}/{\rm d}t_{\rm floor,Alfven})^2}$, 
where $|\Delta x|_{\rm min}$ is the smallest of the cell's sizes along 
$r$, $\theta$, and $\phi$ directions, and d$t_{\rm floor,Alfven}$ 
is set to $3\times10^5$~seconds. As a result, artificial mass 
is added to cells with densities below the minimum density. Such a floor 
usually triggers in the bipolar regions with very large Alfv{\'e}n 
speeds and often very tiny $|\Delta x|_\phi$.}
to avoid extremely small timesteps, which may artificially add 
tiny mass to the outflow regions and can alter the momentum 
of the outflow. Therefore, the discussion in this section is 
more qualitative than quantitative.
\begin{figure*}
\includegraphics[width=1.0\textwidth,left]{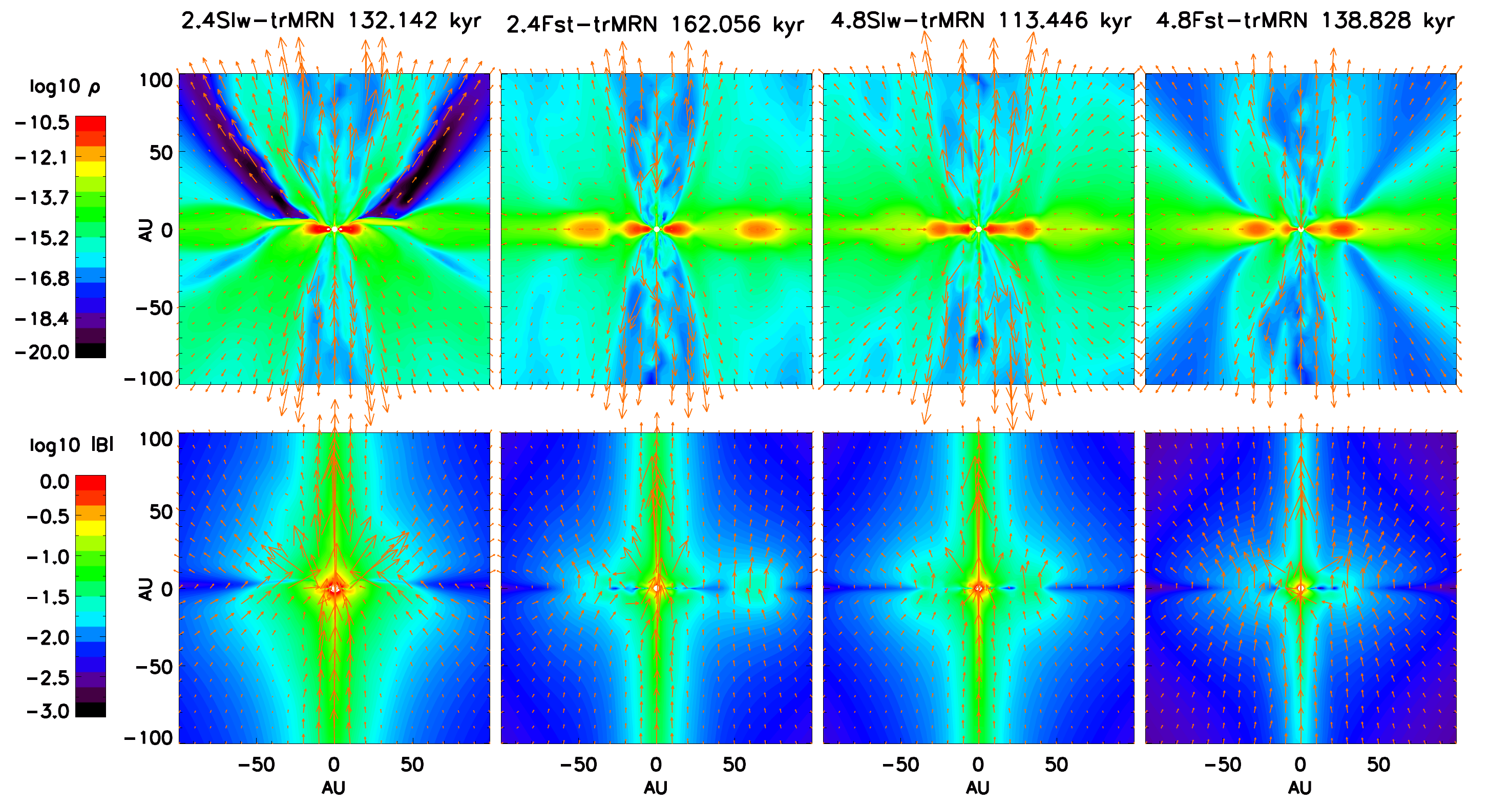}
\caption{Comparison of outflow structures in the bipolar region 
(top row) and magnetic field configurations (bottom row) 
among the disc forming models. The colour contours are logarithmic 
mass density $\rho$ (g~cm$^{-3}$) and magnetic field strength |$B$| (G) 
for top and bottom rows, respectively. The red arrows are 
velocity field and magnetic field vectors on the meridian plane 
for the top and bottom rows, respectively.}
\label{Fig:Outflow_comp}
\end{figure*}

As shown in Fig.~\ref{Fig:Outflow_comp}, the outflow in the 2.4Slw-trMRN 
model has two components: one is centrally collimated and the other 
fanning out sideways. The former bipolar component may correspond to 
the jet or ``X-wind'' type \citep{Shu+2000}; yet simulations with 
much higher resolution are required to understand its nature, which 
is beyond the scope of this study. The latter fan-like component, 
however, only appears in the upper hemisphere (similar to 
the 4.8Fst-MRN model). The origin of this outflow component is 
likely to be the magneto-centrifugal ('slingshot') mechanism 
\citep{BlandfordPayne1982}, in that its launching point coincides 
with the landing site of the 'curved' infall stream, where magnetic 
field lines are strongly pinched. The bottom row of the 1st panel 
from the left clearly 
shows that the pinching locations of magnetic field lines are 
shifted towards the upper hemisphere, following the curved infall 
stream. Therefore, the pseudo-disc is also curved instead of a plane 
along the equator. Note that the edge-on picture is very similar 
to the shrinking disc model (4.8Fst-MRN) discussed above, except that 
the landing site of the infall stream is at $\sim$10~AU compared with 
$\lesssim$3~AU in the shrinking disc model, thanks to the high 
angular momentum influx in the absence of VSGs.

In the other three tr-MRN models, the outer fan-like component 
of the outflow is not as prominent as the 2.4Slw-trMRN model; 
while the inner collimated component is clearly visible. 
The reason can be numerical (Alfv{\'e}n d$t$ floor) as well as 
physical. As shown in the bottom rows of 2nd--4th panels from the left, 
magnetic field lines indeed pile up at the centrifugal barrier, 
which is in the form of spiral (2nd--3rd panels from the left) 
or ring (4th panel from the left) structures (see corresponding 
frames in Fig.~\ref{Fig:Spiral_evol}--\ref{Fig:Ring_frag}). 
However, the ``ring type'' centrifugal barrier show more obvious 
fan-like outflow cavities than the ``spiral type'' centrifugal 
barrier; this is because the former barrier is more efficient in 
blocking infall and piling up field lines, while the spiral structure 
still allow materials and field lines to further fall and spiral along 
it. Nevertheless, the fan-like outflow should be visible if the 
accretion flow from the envelope lands on a narrow region 
of the disc or ring, which can be an explanation to the outflow offset 
recently observed by \citet{Bjerkeli+2016} and \citet{Alves+2017}. 
In this paradigm, asymmetric (one-sided) outflows can be more 
common than symmetric ones.

\section{Discussion}
\label{Chap.Discuss}

\subsection{DEMS \& AD Shock}

The main difference of our work with previous studies is the relatively
efficient decoupling of magnetic field in the collapsing envelope. 
It not only promotes the formation of RSDs, but also avoids the 
abrupt decoupling of the large amount of magnetic flux dragged into 
the stellar vicinity. Otherwise, as we have demonstrated in models 
with MRN grains, sooner or later, DEMS will dominate the inner 
tens to hundreds of AU regions, obstructing the gas rotation further. 
Similarly, if efficient AD only occurs in the inner tens of AU, 
the ``so-called'' AD-shock \citep{LiMcKee1996} is also inevitable. 
Therefore, the magnetic flux problem and magnetic braking catastrophe 
in star formation are closely related. Averting the former will 
simultaneously alleviate the latter, which can be achieved by 
the removal of VSGs in the collapsing protostellar envelope 
or even earlier, during the prestellar core.

\subsection{Disc Size}

Among the tr-MRN models, the inner RSDs (or circumstellar discs) at later 
times are of $\sim$20~AU in radius and are surrounded by the extended 
spiral structures. This picture seemingly matches the self-regulated disc 
formation model by \citet{Hennebelle+2016}. However, even the 
non-magnetic models (Table~\ref{Tab:models}) produce small circumstellar 
discs with $\sim$20~AU radius; yet their rotation-dominated spiral 
structures are much larger. Hence, the small disc radius is unlikely to 
result from magnetic effects. In fact, the size of the circumstellar disc 
is mainly determined by the thermal support (Fig.~\ref{Fig:HydroSlw}) 
and is sensitive to the adiabatic index (see Appendix~\ref{App.A}). 
In comparison, the extended spiral structures are as thick as 
100--150~AU in the vertical direction and are not in hydrostatic 
equilibrium at large scale heights; the outer layers are largely 
infalling towards the midplane. Therefore, the extended spiral 
structure may not be strictly defined as part of the disc, but 
a transition zone between the envelope and the circumstellar disc, 
where infalling gas starts to spiral inward.
Furthermore, when considering the magnetic flux distribution during 
the collapse, the dominant term should be ${\partial B_r \over \partial z}$, 
i.e., field pinching along the infall plane as we have shown in 
\S.~\ref{S.fieldPinch}, instead of $\partial B_z \over \partial r$ 
considered in \citet{Hennebelle+2016}. 
\begin{figure}
\includegraphics[width=1.0\columnwidth,left]{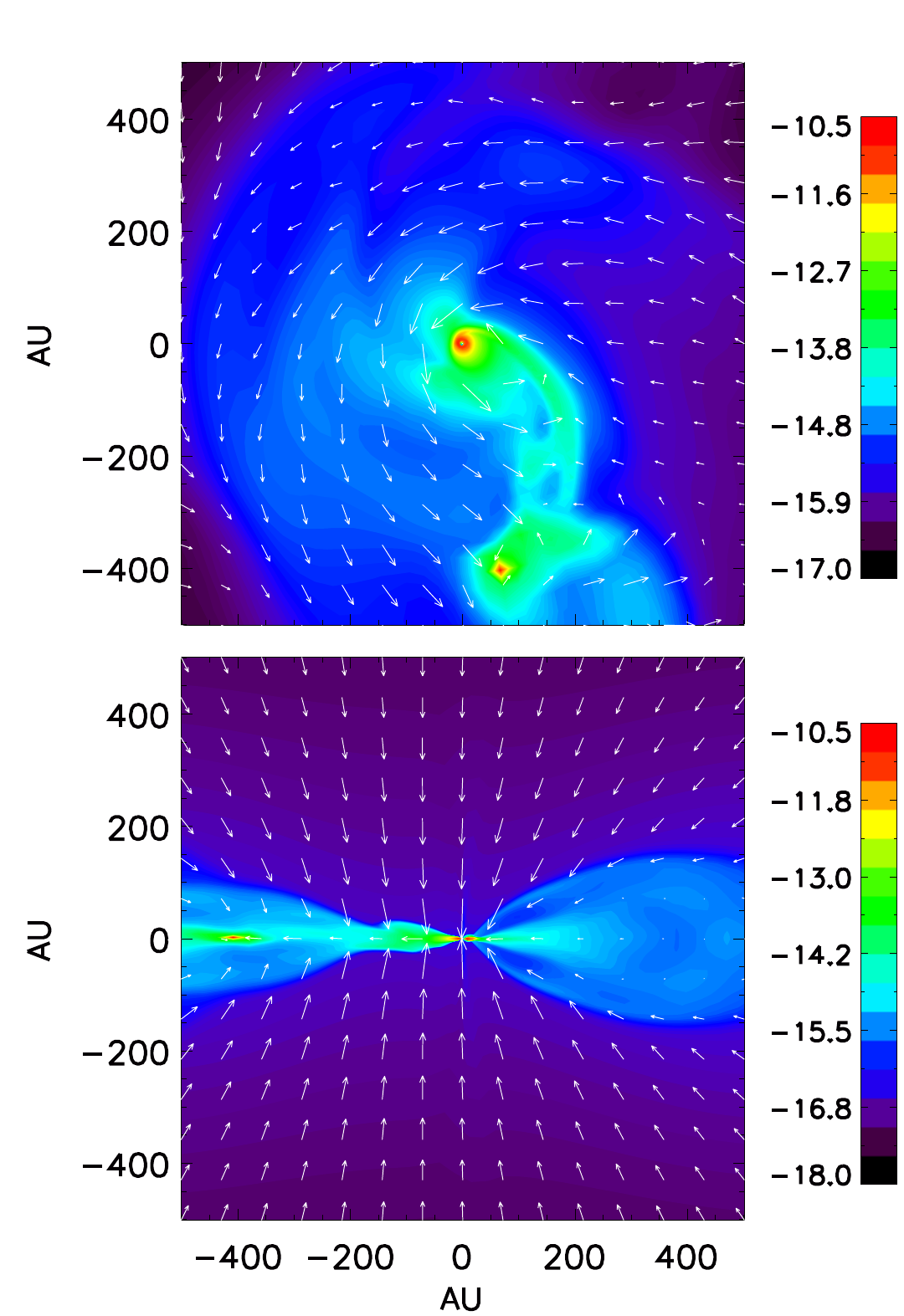}
\caption{Binary system formed in the HydroSlw model at $t\approx119.5$~kyr, 
with the central object of 0.087~M$_{\sun}$. Both circumstellar 
discs are $\sim$10--20~AU in radius. The bulk part of the circumstellar disc 
has a temperature of 50--100~K. The distribution of logarithmic mass density 
and vectors of velocity fields are plotted within 0.336$^{\circ}$ of the 
equator (top) and on the meridian plane (bottom).}
\label{Fig:HydroSlw}
\end{figure}

The main difference between the non-magnetic and magnetic models 
is the size of the spiral structure, which is 
much larger in the non-magnetic models (Table~\ref{Tab:models}). 
The spiral structure marks the regions where rotation starts to 
dominate over infall (hints of large spiral structures in recent 
observations: \ct{Tokuda+2014}, \ct{Perez+2016}, and \ct{Yen+2017}). 
Thus, the outer edge of the spiral structure is generally the 
location where the specific angular momentum curve starts to 
flatten \citep[e.g.,]{Goodman+1993,Belloche2013}, 
or the plateau of the centrifugal radius curve. Because of the 
quickening of rotational motion, the magnetic braking is also 
the strongest inside the spiral regions (e.g., Fig.~\ref{Fig:Torq_comp}), 
which makes the spiral structure smaller over time in models with 
stronger magnetic field.

\subsection{Lower Limit for \texorpdfstring{$\beta_{\rm rot}$} \\
\& The Case of B335}

As we have shown that both initial rotation and magnetization can 
affect the formation of RSDs, we thus explore the lower limit of 
initial rotation speed below which an RSD is hard to form. From 
the test simulations we carried out for tr-MRN grains,\footnote{
The initial angular speed $\omega_0$ tested includes 
$8\times10^{-14}$~s$^{-1}$ and $7.2\times10^{-14}$~s$^{-1}$ for 
initial magnetization of $\lambda=2.4$, and $6.4\times10^{-14}$~s$^{-1}$, 
$5\times10^{-14}$~s$^{-1}$, and $4\times10^{-14}$~s$^{-1}$ for initial 
magnetization of $\lambda=4.8$.} we find that, in order for discs to 
form, the ratio of rotational to gravitational energy 
$\beta_{\rm rot}$ should be higher than $\sim$1.5--1.6\% for initial 
magnetization of $\lambda$=2.4 and $\sim$0.5--0.6\% for magnetization 
of $\lambda$=4.8, which corresponds to a lower limit of angular speed 
$\omega_0$=$8\times10^{-14}$~s$^{-1}$ and 
$5\times10^{-14}$~s$^{-1}$ respectively. Note that these values 
only apply to dense cores with initial solid-body rotation profile.\footnote{
The lower limit of $\beta_{\rm rot}$ found here is likely to be a key 
for the small disks ($\lesssim$10~AU) formed in \ct{Dapp+2012}, apart from 
their high cosmic-ray ionization rate ($\zeta_0^{\rm H_2}=5.0\times 10^{-17}$~s$^{-1}$). 
In Figure 2 of \ct{Dapp+2012}, they showed very similar trend of enhanced 
AD in the absence of VSGs as we presented in Zhao+16; however, they did not 
elaborate on its effect on disk formation. The initial rotation speed 
used in \citet{Dapp+2012} as well as in \citet{DappBasu2010} 
(1~km~s$^{-1}$~pc$^{-1}\approx3.2\times10^{-14}$~s$^{-1}$) seems to be 
too slow to produce disks larger than 10~AU. It corresponds to 
$\beta_{\rm rot}$$\sim$0.6\% for their $\lambda=2$ core, which is 
well below the lower limit we found for the $\lambda=2.4$ case.}

\begin{figure}
\includegraphics[width=1.0\columnwidth,left]{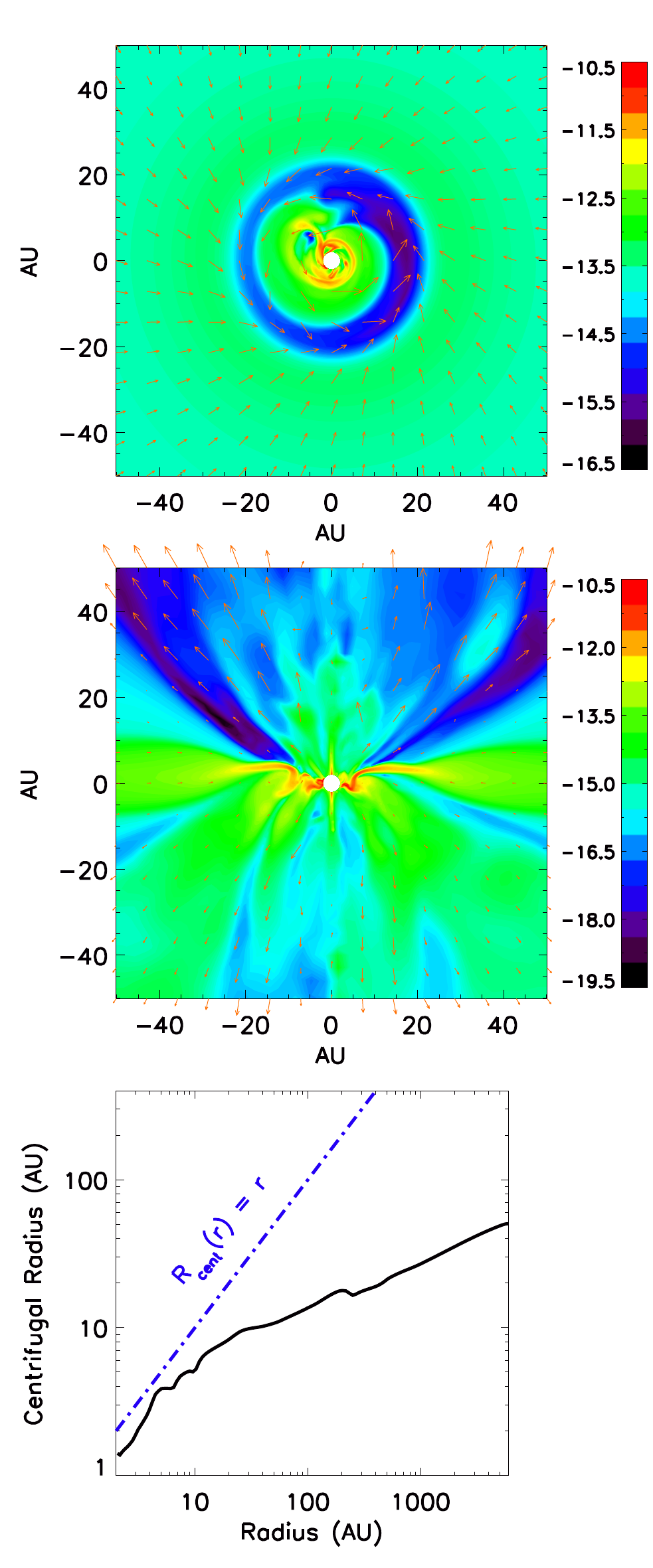}
\caption{Distribution of logarithmic mass density $\rho$ (g~cm$^{-3}$)
within 0.336$^{\circ}$ of the equator (top) and on the meridian plane 
(middle) for a slow rotating core 
($\omega_0$=$4\times10^{-14}$~s$^{-1}$) with $\lambda=4.8$ 
at $t\approx107.9$~kyr. The mass of the central object is 
$\sim$0.06~M$_{\sun}$. Orange vectors are velocity fields. 
The DEMS are constantly disrupting the disc structure (<10~AU) 
in the centre. Outflows are yet prominent. 
Bottom: estimated centrifugal radius.}
\label{Fig:4.8extSlw}
\end{figure}

When the initial rotation is slower than the lower limit, the infalling 
gas will reach a centrifugal radius of <10~AU, dragging a large amount 
of magnetic flux to the stellar vicinity (even under the enhanced AD 
due to removal of VSGs). Therefore, the central $\sim$10~AU region 
is inevitably dominated by DEMS, which constantly disrupt the disc 
structure and suppress the formation of a well-defined RSD. As shown 
in Fig.~\ref{Fig:4.8extSlw}, the central object is surrounded by 
filamentary accretion streams rather than a disc (face-on view). 
In the edge-on view, the infall stream curves towards the upper 
hemisphere and lands onto the disc from above, which results in 
an one-sided outflow (similar to the 2.4Slw-trMRN and 4.8Fst-MRN 
models above\footnote{In some test simulations, the infall stream 
can also curve towards the lower hemisphere and land onto the disc 
from below.}). The curve of centrifugal radius decreases nearly 
monotonically towards the inner 10~AU, only showing a small 
plateau at $\sim$5~AU. It is consistent with the landing radius 
of the infall stream. 

This particular model matches to some extent the observation of 
B335, where no Keplerian disc >10~AU is found around the protostar 
\citep{Yen+2015b,Evans+2015} yet CO outflows are clearly visible 
\citep{Yen+2010}. The estimated angular speed of the host core is 
about $2.6\times10^{-14}$~s$^{-1}$, which is a few times lower than 
other protostellar sources with disc detection \citep{Yen+2015}. 
This value is close to the lower limit of angular speed we find 
for $\lambda=4.8$ cores. It is likely that B335 is already in lack 
of VSGs during the collapse process; otherwise, a much larger 
angular speed (>10$^{-13}$~s$^{-1}$) or mass-to-flux ratio (>10) of 
the host core is required with the full MRN distribution to form even 
a small, short-lived (or shrinking) disc (\S.~\ref{S.shrinkDisc}). 
Nevertheless, a disc, even a small one, is necessary to power 
the ordered, wide-angle outflows observed in B335; 
a fully magnetically dominated inner region with no disc at all 
generally produces chaotic morphologies in the bipolar region, 
which is unlikely the case for B335.

\subsection{VSGs in Dense Cores}

The lack of VSGs in dense cores \citep{Tibbs+2016} can be caused by 
either (1) grain evolution or (2) magnetic selection effect. 
The former includes accretion of VSGs onto grain mantles 
\citep[process analogous to molecular freeze-out, e.g.,][]{TielensHagen1982,Hasegawa+1992} 
and grain coagulation \citep{Chokshi+1993,DominikTielens1997}; 
and the latter refers to the differential coupling of grains to the 
magnetic field based on their size \citep{CiolekMouschovias1996}. 
While grain evolution can start from the quiescent prestellar phase, 
magnetic selection normally occurs during the collapse phase. 
We will leave the more complete chemistry model including grain 
evolution and multi-fluid non-ideal MHD to future studies.

\subsection{Numerical Limitations}

Finally, we raise cautions for the potential limitations of our numerical 
treatment. \\
{\it First}, we use Alfv{\'e}n d$t$ floor to limit the velocity 
of MHD waves, which will add tiny mass to certain grid cells 
with low density but strong magnetic field (the total added mass for 
the entire computational domain is on the order of 
$\sim$10$^{-3}$--10$^{-2}$ of the central point mass). The floor is 
most likely to trigger in the bipolar regions, which can affect the 
outflow dynamics and more specifically, slowing down the outflow. \\
{\it Second}, we set an AD d$t$ floor (d$t_{\rm floor,AD}$) to avoid 
intolerably small AD timesteps. The result is that the AD diffusivity 
$\eta_{\rm AD}$ is capped\footnote{The cap of $\eta_{\rm AD}$ 
is computed for each cell as 
${\rm CFL_{AD}} {|\Delta x|_{\rm min}^2 \over 4~{\rm d}t_{\rm floor,AD}}$, 
where ${\rm CFL_{AD}}$=0.4 is the Courant-Friedrichs-Lewy number for 
AD, $|\Delta x|_{\rm min}$ is the smallest of the cell's sizes along 
$r$, $\theta$ and $\phi$ directions, and d$t_{\rm floor,AD}$ is set 
to $3\times10^6$~seconds. Such an $\eta_{\rm AD}$ cap 
based on d$t_{\rm floor,AD}$ behaves approximately as a constant 
$\eta_{\rm AD}$ cap of $\sim$10$^{19}$~cm$^2$~s$^{-1}$ in the 
bulk part of the circumstellar disc.} 
in the outflow region and bulk part of the disc (if any), which causes 
the matter there to be better coupled to the magnetic fields than it 
should be. This cap on $\eta_{\rm AD}$ should not affect our main 
result that disc formation is enabled by ambipolar diffusion 
in the infalling envelope. We have explored smaller values of 
d$t_{\rm floor,AD}$ (=$1\times10^6$ and $3\times10^5$) using a lower 
resolution in r-direction ($\delta r=0.33$~AU), and find that large 
DEMS still dominate the central regions around the star and disc 
formation is strongly suppressed as in \S~\ref{S.model-MRN}. 
Therefore, in terms of forming RSDs, the specific angular momentum 
of the infalling gas being accreted 
by the disc plays a dominant role over the decoupling process in 
the disc itself. \\
{\it Third}, the code has no sink particle treatment (except for the 
stationary sink hole in the centre), and hence is not 
ideal for following orbital evolution of binary and multiple 
systems. We only confirm the onset of fragmentation and the 
formation of companion stellar masses. \\
{\it Fourth}, we have not included Ohmic and Hall effect in this 
study, but their corresponding diffusivities are computed regardlessly. 
We found in the tr-MRN models the Ohmic diffusivity is about $\sim$10 
times smaller than the ambipolar diffusivity inside the circumstellar 
discs. As we have shown in Zhao+16, Ohmic dissipation has negligible 
effect on disc formation because (1) with the full MRN distribution, 
the lack of dense long-lived disc in the first place makes it 
difficult for Ohmic dissipation to dominate; while (2) in the 
absence of VSGs, ambipolar diffusivity always dominates over 
Ohmic diffusivity at densities below 10$^{14}$--10$^{15}$~cm$^{-3}$. 
Once we further decrease the radius of the inner boundary (e.g. below 1~AU), 
the inclusion of Ohmic dissipation will become crucial. Besides, 
Hall effect is very sensitive to the polarity of magnetic field 
\citep{Krasnopolsky+2011,BraidingWardle2012,Tsukamoto+2015b,Wurster+2016}, 
which adds another twist to the formation of protostellar discs. 
We will investigate the interplay between the three non-ideal 
MHD effects in future studies.

\section{Summary}
\label{Chap.Summary}

We have extended our 2D study of protostellar disc formation (Zhao+16) 
to 3D, using the same equilibrium chemical network. We have verified 
the main result found in Zhao+16 that the removal of VSGs enables 
the formation of RSDs of tens of AU through the enhanced ambipolar 
diffusion of magnetic fields in the collapsing envelope. Because of 
the reduction of magnetic flux dragged into the central disc 
forming region, magnetic braking becomes less efficient and the 
key ingredient for disc formation --- {\it angular momentum} --- 
is sufficiently retained. Further conclusions are listed below. 
\begin{description}
\item 1. Collapse pinches magnetic field lines along the infall 
plane (pseudo-disc plane) where ambipolar drift is the strongest 
($\propto {\partial B_r \over \partial \theta}$). With the 
enhanced $\eta_{\rm AD}$ (VSGs absent), the ``effective'' ion velocity 
(or the magnetic field velocity) nearly vanishes along the infall 
plane even at 10$^2$--10$^3$~AU scales. This indicates that magnetic 
fields are gradually left behind in the envelope by the collapsing 
neutral matter. 
\item 2. The so-called DEMS (formed by the decoupled magnetic flux 
from the accreted matter) can still suppress disc formation and 
obstruct disc rotation if the full MRN distribution is used. 
Even in the absence of VSGs, DEMS are still present at early times. 
They play the role of disrupting the first core, preventing it from 
directly evolving into a circumstellar disc. Disc formation in such 
cases relies on the supply of high angular momentum material from 
the infalling envelope. 
\item 3. High specific angular momentum entering the inner region 
enables both the formation of RSDs and the development 
of spiral (or ring) structures. The size of spirals (or rings) 
matches the centrifugal radius of the infalling gas. The outer edge 
of the spirals (or rings) is also the location where the 
specific angular momentum curve starts to flatten. 
\item 4. When the initial core is relatively weakly magnetized 
(with a dimensionless mass-to-flux ratio of 4.8 rather than 2.4), 
fragmentation frequently occurs on the outer spiral arms 
(or on the rings), where infall material from the envelope 
piles up. Such fragments of tens of Jupiter mass could mark the 
early onset of multiple systems. Some of the fragments, particularly 
when the core rotation is relatively slow, merge quickly with 
the central stellar object, producing bursts of mass accretion. 
Others are longer lived, and can potentially survive as multiple 
stellar systems, although sink particles are needed to better 
follow their long-term evolution. 
\item 5. Magnetically-driven outflows have multiple components. 
The central bipolar component is more symmetric and collimated, 
which may correspond to jet or ``X-wind''. The outer fan-like 
component can be more asymmetric (one-sided). Its launching 
location coincides with the landing site of the infalling gas 
on the disc or ring. 
\end{description}

\section*{Acknowledgements}

We thank Kengo Tomida for providing the data of evolutionary track. 
BZ and PC acknowledge support from the European Research Council 
(ERC; project PALs 320620). 
Z.-Y. L. is supported in part by NASA NNX14AB38G, 
NSF AST-1313083, and NSF AST-1716259. 
Numerical simulations are carried out on the CAS group cluster at MPE.


\appendix

\section{Equation of State}
\label{App.A}

\begin{figure*}
\includegraphics[width=\textwidth]{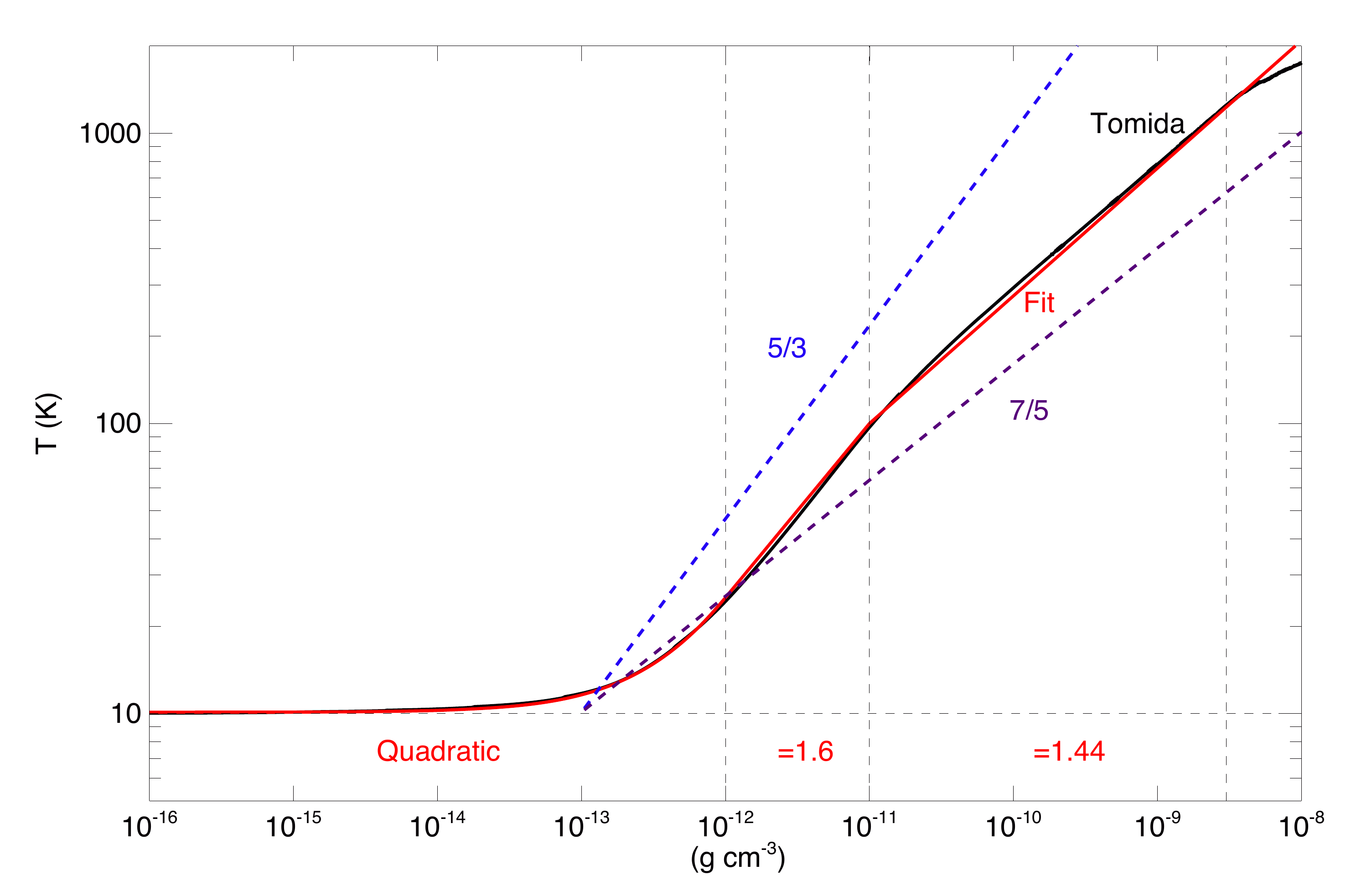}
\caption{Function fitting (red solid curve) for the evolutionary track 
from \citet{Tomida+2013} (black solid curve). The barotropic EOS 
with adiabatic indices of 5/3 and 7/5 are shown in blue dashed and 
purple dashed lines.}
\label{Fig:EOS}
\end{figure*}

We use a broken power law profile for the equation of state (EOS), 
fitted to mimic the radiative transfer results of \citet{Tomida+2013}: 
\begin{equation}
T= \left\{\begin{array}{lcl} T_0 + 1.5{\rho \over 10^{-13}} & \mbox{for} & \rho<10^{-12} \\
(T_0+15)({\rho \over 10^{-12}})^{0.6} & \mbox{for} & 10^{-12} \leqslant \rho \leqslant 10^{-11} \\
10^{0.6}(T_0+15)({\rho \over 10^{-11}})^{0.44} & \mbox{for} & 10^{-11} \leqslant \rho \leqslant 3\times10^{-9} \end{array}\right.
\end{equation}
where T$_0 = 10$~K. The comparison for different EOS is shown in 
Fig.~\ref{Fig:EOS}. The conventional 5/3 law reproduces the right slope below 
10$^{-11}$~g~cm$^{-3}$ but overestimates the gas pressure by a factor of 
$\sim$2, while the 7/5 law reproduces the right slope beyond 
10$^{-11}$~g~cm$^{-3}$ but underestimates the gas pressure by a factor of 
$\sim$2. As a result, the usual 5/3 law tends to form thermally 
supported structures with somewhat larger sizes, while the structures formed 
by the 7/5 law are smaller and vulnerable to gravitationally compression. 
Note that the density regime which matters the most for disc formation is 
below 10$^{-11}$~g~cm$^{-3}$ \citep{Zhao+2016}, hence the 5/3 power law 
provides better thermal stability than the 7/5 profile.\footnote{We find that 
stiffening the EOS at $3\times10^{-13}$~g~cm$^{-3}$ offers a more realistic 
thermal pressure than at $1\times10^{-13}$~g~cm$^{-3}$ for the single 
5/3 power law.}


\bsp	
\label{lastpage}
\end{document}